# Past and Present Dynamics of the Iron Biogeochemical Cycle

Nicolas Dauphas[1], Andy W. Heard[2], Eric Siciliano Rego[1], Olivier Rouxel[3], Johanna Marin-Carbonne[4], Virgil Pasquier[4], Andrey Bekker[5], David Rowley[1]

[1] Origins Laboratory, Department of the Geophysical Sciences and Enrico Fermi Institute, The University of Chicago, 5734 South Ellis, Chicago IL 60637, USA
[2] Department of Geology & Geophysics, Woods Hole Oceanographic Institution, Woods Hole, MA, USA
[3] IFREMER, Centre de Brest, Geo-Ocean UMR 6538, F-29280 Plouzané, France
[4] Institut des Sciences de la Terre, Université de Lausanne, Lausanne, 1015, Switzerland
[5] Department of Earth & Planetary Sciences, University of California, Riverside, CA 92521, USA

**Keypoints/objectives:**
- Quantify source and sink fluxes of dissolved iron in the Archean, Proterozoic, and Present
- Evaluate how the iron biogeochemical cycle in the oceans was affected by Earth's oxygenation
- Quantify how iron's residence time in the oceans changed through time
- Identify knowledge gaps in our understanding of the iron cycle

**Abstract**

**This chapter investigates the complexities surrounding the iron biogeochemical cycle from the Archean to present, with a focus on assessing the balance between iron sources and sinks during long periods of Earth's history with relatively invariable redox conditions, when steady state can be safely assumed. Currently, the residence time of iron in the ocean may be as short as approximately 5 years. The input flux of iron is highly sensitive to redox cycling in sediments, while its removal primarily occurs through dispersed processes of oxidation and precipitation. In the Archean, we find a significant imbalance between continental and hydrothermal inputs, which collectively contribute between 61,500 to 263,000 $Gg.yr^{-1}$ of dissolved iron to the oceans, and the most obvious sinks such as iron formations (IFs), which sequester up to ~43,000 $Gg.yr^{-1}$ of iron. A possible solution to this imbalance involves the dispersed abiotic precipitation and removal of iron as silicates, sulfides, and carbonates in marine basins. Additionally, we calculate the residence time of dissolved iron in the Archean oceans to be between 6 kyr and 3 Myr, which is significantly longer than the ocean mixing timescale. Our estimates indicate that under the anoxic Archean atmosphere, the iron cycle was more protracted than today, and the isotopic compositions and concentrations of dissolved iron were likely more uniform. Distinct water bodies were likely confined to limited areas or specific, dynamic systems with intense iron turnover, such as regions where deep-sea upwelling currents brought hydrothermal iron to photic zones rich in biotic or abiotic oxidants.**

# 1. Introduction

Planetary objects are replete with iron, but this element is often permanently locked in their metallic cores. It is estimated that only 13% of Earth's inventory of iron is presently in the mantle and crust (McDonough and Sun, 1995). Most iron in the upper mantle is present as ferrous $Fe^{2+}$ (~96.2%), while the rest exists as ferric $Fe^{3+}$ (~3.8%) (Canil et al., 1994; Rudra and Hirschmann, 2022). Some metallic iron exists in the mantle where it can form through the disproportionation reaction $3Fe^{2+} \rightarrow Fe^{0} + 2Fe^{3+}$ (Frost et al., 2004; Rohrbach et al., 2007; Smith et al., 2016). Terrestrial surface environments in contact with the oxygenated atmosphere contain iron primarily as $Fe^{3+}$, often as sedimentary oxyhydroxide minerals. The redox state of iron on Earth is thus stratified and is marked by a large disequilibrium between oxygenated surface and reduced interior conditions. On long geological timescales, iron can be cycled between Earth's interior and surface. It is delivered to the crust from the mantle with magmatism, and is transferred from the surface back to the mantle through recycling at subduction zones or crustal delamination. The oceanic iron cycle operates on a shorter timescale and sedimentary archives bear witness to changes in the biogeochemical cycle of iron through eons. Through redox reactions, the global iron biogeochemical cycle is tied to the cycles of other bio-essential redox elements carbon and sulfur. To tell iron's story is also thus also to tell the story of the development of life and Earth's oxygenation (Holland, 1984; Lee et al., 2016).

In the present contribution, we examine how the dissolved iron biogeochemical (DIB) cycle in the oceans could have evolved as the Earth transitioned from a predominantly anoxic state in the Archean to an almost fully oxic one at present. Our approach in tackling these changes is to quantify the iron flux of all iron sources and sinks that could have contributed to that cycle. We caution the reader that this exercise is fraught with difficulties as even at present, considerable uncertainties remain to close the iron cycle (Boyd and Ellwood, 2010; Raiswell and Canfield, 2012), with controls often happening at scales and in settings that are seldom recorded in geological archives. With this caveat in mind, we see this as a valuable effort as it reveals where knowledge gaps are present, helping shape future efforts to improve on our understanding of the drivers of biological productivity and Earth's oxygenation.

The modern DIB cycle has been extensively studied due to iron's rate-limiting role in biological productivity in certain parts of the oceans, such as the Southern Ocean (Boyd et al., 2000; Martin, 1990; Moore et al., 2013). The modern iron cycle discussed in Sect. 2 provides the framework for interpreting iron behavior on Earth in the past if sources and sinks changed in flux but not in nature. Paradoxically, some aspects of the iron cycle in ancient oceans are better constrained than in modern oceans because iron was less affected by redox and biochemical processes under more reducing conditions. The sizes of the dissolved iron fluxes in and out of the oceans (Fig. 1) changed in response to changes in environmental conditions and evolutionary innovations in metabolic processes involving iron. The different iron fluxes associated with continental export, hydrothermal systems, and sedimentary sources and sinks are discussed and quantified in sections 3, 4, and 5. Each term of the DIB cycle was handled independently by one or several authors. In section 6, we compare sources and sinks and revise them if needed to bring the cycle to closure.

Sections 3 and onwards examine secular changes in DIB sources and sinks by dividing Earth's history into several time periods commonly adopted for describing the main stages of Earth's atmospheric oxygenation (Holland, 2006). Before delving into the details of the DIB cycle, we touch below on several aspects that concern all sections. We are primarily interested in dissolved iron, which is involved in biochemical reactions and is the source of chemical sedimentary rocks and authigenic minerals. An operative definition of dissolved iron is needed as there is a continuum of sizes between truly dissolved and particulate iron, and nanoparticles can have long residence times and be very reactive. We build our understanding of the past iron cycle based on the knowledge of the modern iron cycle. Modern waters are usually filtered to separate a suspended and filterable fraction, with a cutoff between the two at around 0.45 μm. In discussing the modern DIB cycle, we therefore take the <0.45 μm filterable fraction as proxy for dissolved iron.

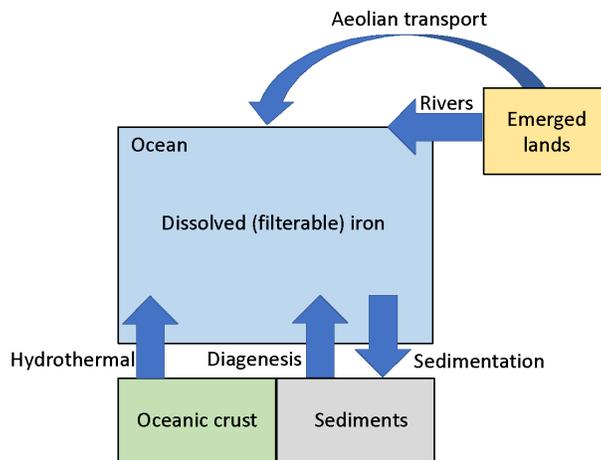

**Fig. 1.** Fluxes considered in the dissolved iron biogeochemical (DIB) cycle through time.

Iron surface cycling saw dramatic changes associated with two major episodes of atmospheric oxygenation at the Great Oxidation Event (GOE) and Neoproterozoic Oxidation Event (NOE) (Bekker and Holland, 2012; Lyons et al., 2014). The beginning of the GOE is usually taken to be marked by the disappearance of sulfur mass-independent fractionation (MIF) (Farquhar et al., 2000) at ~2.43-2.22 Ga (Luo et al., 2016; Gumsley et al., 2017; Warke et al., 2020; Bekker et al., 2004, 2020; Poulton et al., 2021), with the caveat that atmospheric oxygen levels could have fluctuated up and down until ~2.22 Ga and the onset of the Lomagundi-Jatuli carbon isotope excursion (Poulton et al., 2021). Geological archives indicate cessation of widespread banded iron formation deposition by ~2.43 Ga (Bekker et al., 2010), but significant biogeochemical rearrangements involving the iron-sulfur cycle were still taking place until ~1.88 Ga, notably deposition of granular iron formations (Rasmussen et al., 2012; Johnson and Molnar, 2019) associated with the emplacement of Large Igneous Provinces on multiple cratons. Following Holland (2006), we take the GOE as covering the period ~2.4 to ~2.1 Ga. The NOE is associated with an increase in the variability of the carbon isotopic composition of carbonates starting at ~0.8 Ga (Turner and Bekker, 2016; Kuznetsov et al., 2017; Cramer and Jarvis, 2020; Halverson et al., 2005; Knoll et al., 1986), which marks the beginning of the NOE. The end of the NOE coincides approximately with the first fossil evidence for animal locomotion at ~0.56 Ga (Liu et al., 2010), which requires bottom-water oxygenation. These dramatic oxidation episodes (the GOE and NOE) alternated with long periods of stasis that did not see much change in the composition of the oceans and biogeochemical cycles. Knowledge about transient oxidation events is sparse, mostly because their time span is limited, and the rock archives are incomplete. The focus of this review is therefore on the long periods of time when the redox conditions were stable, but we discuss the transient oxidation events when relevant. A virtue of focusing on these stable periods of Earth's history is that we can reasonably assume that dissolved iron content in the oceans was at nearly steady state, meaning that the sources and sinks were well balanced.

We divide Earth's oxygenation history into five stages (Fig. 2) (Holland, 2006): pre-GOE (3.85-2.4 Ga; stage #1), GOE (2.4-2.1 Ga, stage #2), post-GOE to pre-NOE (2.1-0.8 Ga; stage #3), NOE (0.8-0.56 Ga; stage #4), and post-NOE to present (0.56-0 Ga; stage #5). The periods of near stasis in Earth's redox history are stages #1, #3, and #5.

- Pre-GOE (stage #1), the presence of detrital grains of pyrite and uraninite (Grandstaff, 1980; Johnson et al., 2014), mass-independent fractionation (MIF) for sulfur isotopes (Farquhar et al., 2000; Pavlov and Kasting, 2002), and iron mobility in paleosols (Rye and Holland, 1998) all point to an atmosphere that was globally devoid of oxygen (Cloud, 1968, 1972; Holland, 2020; Lyons et al., 2014). An upper limit on Archean atmospheric $PO_2$ of $<10^{-6}$ PAL (Present Atmospheric Level) is given by sulfur MIF (Pavlov and Kasting, 2002; Catling and Zahnle, 2020). Molybdenum and sulfur enrichments in shales point to some degree of oxidative weathering of sulfide through most of the Archean, corresponding to a lower limit on $PO_2$ of $\gtrsim 10^{-7}$ PAL if oxidation was mediated by $O_2$ (Stüeken et al., 2012; Johnson et al., 2021). Global $PO_2$ possibly reached higher levels in the ramp up towards the GOE, which is manifested as large Mo enrichments in shales that are known as oxygen whiffs (Anbar et al., 2007; Ostrander et al., 2021; however see Slotznick et al., 2022 for an alternative metasomatic interpretation; and Anbar et al., 2023 and Slotznick et al., 2023 for the following discussion). While Mo enrichments likely point to pyrite weathering, it may not necessarily have been mediated by free $O_2$, but rather by $Fe^{3+}$ produced by photooxidation (Braterman et al., 1983; Nie et al., 2017), in which case Mo enrichments in shales may not reflect the oxygenation status of the atmosphere, but could instead reflect emergence of the continents and exposure of land to sunlight (Hao et al., 2022). The presence of extensive iron oxide deposits (iron formations; IFs) also indicates the existence of oxidizing conditions and perhaps even oxygen oases in the photic zone (Bekker et al., 2010; Cloud, 1968; Morris, 1993). The argument has been made that iron oxidation in IFs may be an overprint and that they may have deposited as reduced greenalite or other $Fe^{2+}$-bearing precursor (Heard et al., 2023; Rasmussen et al., 2021). While some iron could have precipitated in this manner, the broad negative correlation between Mn/Fe and $\delta^{56}Fe$ in many of the most well-studied iron formations points to iron oxidation and a resultant decoupling of $Fe^{2+}$ and $Mn^{2+}$ in the water column (Heard et al., 2022; Hiebert et al., 2018; Thibon et al., 2019).

- Starting at around 2.5-2.4 Ga, indicators of global surface anoxia started to disappear, marking a rise in atmospheric oxygen. This was followed by ~500 Myr of global reorganization of biogeochemical cycles associated with large carbon isotope excursions known as the Lomagundi-Jatuli and Shunga-Francevillian events (Bekker, 2022; Karhu and Holland, 1996; Kump et al., 2011; Schidlowski et al., 1976). The system settled to another stable state spanning the period 2.06 to 0.8 Ga (stage #2) that lasted until the NOE and is known as the 'boring billion'. The $pO_2$ of the atmosphere during that stage was intermediate between pre-GOE and modern values, with estimates of 0.001 to 0.1 PAL proposed based on Cr and Fe isotope proxy approaches associated with their own significant uncertainties (Canfield et al., 2018; Cole et al., 2016; Planavsky et al., 2014; Wang et al., 2022; Zhang et al., 2016). While the atmosphere and upper layer of the oceans were likely weakly oxygenated, considerable uncertainty remains regarding the deep ocean and whether or not it was euxinic (anoxic and sulfide-rich), ferruginous (anoxic and iron-rich) (Canfield, 1998; Planavsky et al., 2018; Poulton et al., 2010; Reinhard et al., 2013), or even suboxic (Slack et al., 2007, 2009).

- The end of the Neoproterozoic saw another major biogeochemical reorganization that led to another rise in atmospheric O$_2$ to a level close to that in the modern atmosphere (stage #5), although the coeval deep oxygenation of the global oceans at this precise time point requires careful reexamination (reviewed by Ostrander, 2023). This stage of Earth's history saw significant fluctuations in the $p$O$_2$ of the atmosphere, but its values remained within a factor of 2 of the PAL for much of the Phanerozoic (Berner, 1999; Glasspool and Scott, 2010). The modern world exemplifies how the DIB cycle operated during this Eon. Today, almost all parts of Earth's oceans are oxygenated and anoxic bottom waters only represent 0.2% of the surface coverage of the world's oceans (Bertine and Turekian, 1973; Tissot and Dauphas, 2015; Veeh, 1967).

Our understanding of the DIB cycle stems from analysis of sedimentary and igneous archives, which seldom sample deep-ocean basins. This is an important and unavoidable source of uncertainty when discussing the ancient rock record. Some isotopic proxies have been developed (*e.g.*, Mo, U, Tl, and V) that can potentially help constrain the balance of euxinic, anoxic, suboxic, and oxic sedimentary sinks at a global scale in the oceans by examining ancient shale and carbonate sedimentary archives. Application of these proxies to very ancient rocks however suffers from several shortcomings, including the difficulty in translating isotopic data into real world estimates of surface coverage of various redox sinks (Chen et al., 2021; Reinhard et al., 2013), and the fidelity of sedimentary archives of the isotopic composition of ancient seawater. With this caveat in mind, we examine below the fluxes of the iron sinks and sources through time during stasis stages #1, #3, and #5.

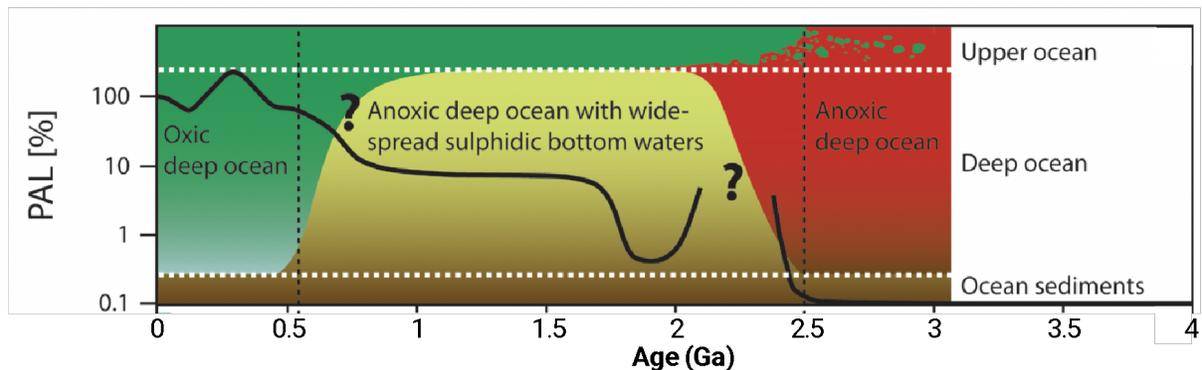

**Fig. 2.** Oxygenation of the atmosphere and oceans through time (modified from Shields-Zhou and Och, 2011).

## 2. Modern ocean iron cycle

Iron in the modern oceans is a trace (< 1 nM) constituent that acts as a critical, and often limiting, micronutrient for marine phytoplankton (Martin et al., 1990a; Martin and Fitzwater, 1988; Martin and Gordon, 1988). In fact, some 25% of the world oceans, namely the Southern Ocean, subarctic Northeast Pacific, and equatorial Pacific, exist as high-nutrient, low-chlorophyll (HNLC; Fig. 3) regions where sufficient macronutrients (N, P) are present for marine primary production, but a dearth of available Fe in the photic zone throttles phytoplankton growth (Martin et al., 1994, 1990b, 1989; Martin and Fitzwater, 1988; Moore et al., 2001). Iron limitation of primary productivity in HNLC regions was demonstrated during large-scale Fe fertilization experiments in the equatorial Pacific in the mid-1990s (Coale et al., 1996b; Martin et al., 1994).

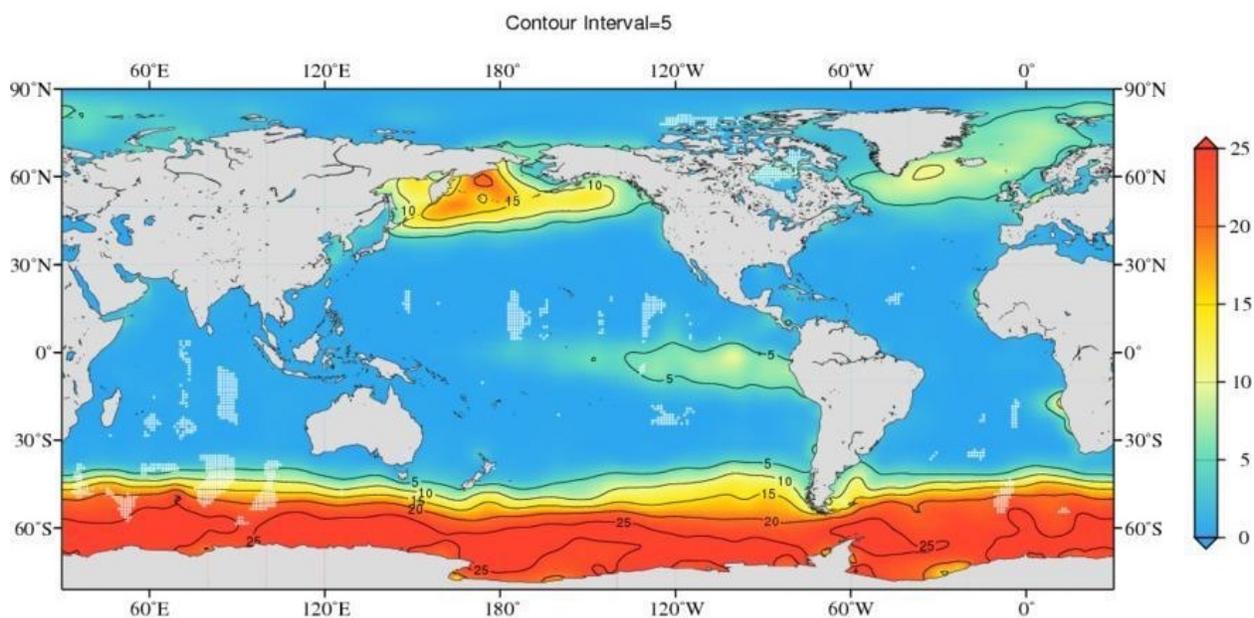

**Fig. 3.** Annual nitrate surface concentrations in the oceans (Garcia et al., 2019). The Southern Ocean, Equatorial Pacific, and Subarctic North Pacific contain large nitrate concentrations, but primary productivity is low. In those regions known as high-nutrient-low-chlorophyll (HNLC) areas, primary productivity is limited by iron.

The low concentration of dissolved Fe in the modern global oceans, even in comparison to transition elements that are orders of magnitude less abundant on Earth, is a consequence of well-oxygenated conditions in Earth surface reservoirs under which the stable form of Fe is highly insoluble, oxidized $Fe^{3+}$ (Millero, 1998). The residence time of Fe in the modern oceans is estimated to be on the order of 200 years (Boyle, 1997; Johnson et al., 1997a), although widely varying estimates of input and output fluxes of Fe to the oceans, and the complex nature of the standing dissolved Fe pool lead to significant uncertainties in calculating this value, with estimates ranging between ~4 and 630 years (Tagliabue et al., 2016).

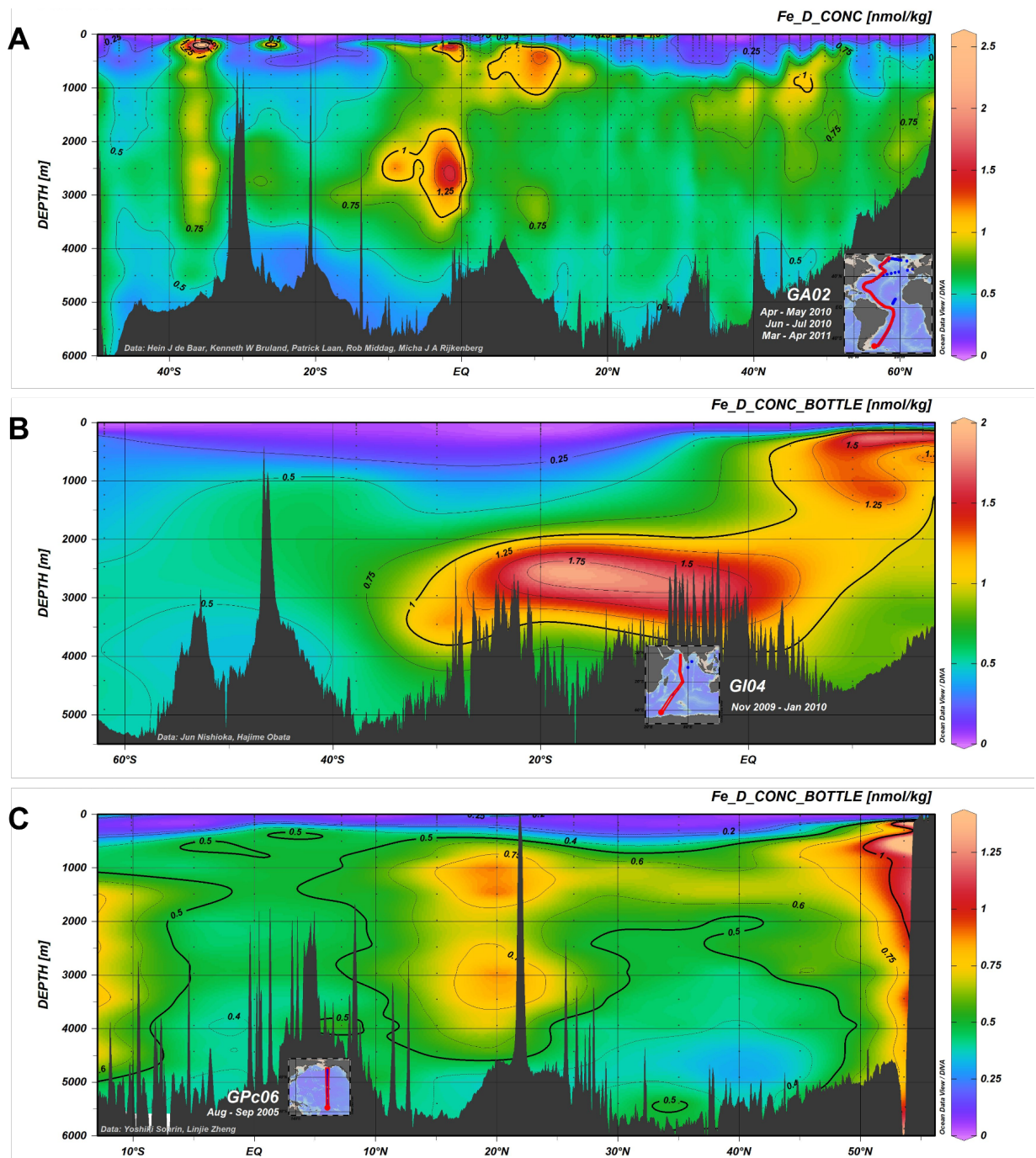

**Fig. 4.** Profiles of dissolved Fe in the modern ocean basins from the GEOTRACES program, showing contours of dissolved Fe in nM. A: The West Atlantic Ocean (GA02; Rijkenberg et al., 2014). B: The Indian Ocean (GI 04; Nishioka et al., 2013; Thi Dieu Vu and Sohrin, 2013). C: The North Pacific Ocean (GPc06; Zheng and Sohrin, 2019).

The surface distribution of Fe in the oceans largely reflects surficial inputs of Fe-bearing aerosol dust (Boyle, 1997; Jickells et al., 2005; Luther and Wu, 1997; Mahowald et al., 2005). High quality measurements of Fe concentration profiles throughout the global oceans have been available since the 1980s (Fig. 4; Bruland et al., 1994; Johnson et al., 1997a; Landing and Bruland, 1987; Martin et al., 1990b, 1989; Martin and Gordon, 1988; Wu and Luther III, 1994) and indicate a generally nutrient-like concentration profile for Fe. Concentrations in the near-surface ocean are 0.02-0.25 nM away from high dust deposition regions, and increase with depth to a value of approximately 0.6 nM beneath the chemocline that is surprisingly consistent throughout large portions of the global oceans (Bruland et al., 1994; Conway and John, 2014; Johnson et al., 1997a; Lacan et al., 2008; Landing and Bruland, 1987; Moore and Braucher, 2008), although by no means as uniform as initially thought (Boyle, 1997; Johnson et al., 1997a; Luther and Wu, 1997; Sunda, 1997). The profile of Fe is distinct from the particle-reactive, scavenging-influenced, element profiles of Al, Mn, and Pb that decrease with depth due to aging of water masses (Bruland et al., 1994; Landing and Bruland, 1987; Orians and Bruland, 1985). Unlike those elements, substantial Fe in phytoplankton is re-mineralized at depth and released back into solution.

The behavior of Fe in the oceans does not entirely follow that of other micronutrient metals like Zn and Cd, either in its depth profile (Bruland et al., 1994), or in its lack of inter-ocean fractionation in comparison to other nutrients that accumulate around the deep-ocean conveyor belt from the North Atlantic to the Pacific (Johnson et al., 1997a). Greater lateral variability in dissolved Fe exists within ocean basins, with elevated concentrations near continental margins. Dissolved Fe can exceed 1 nM in nearshore environments, increasing to several nM in oxygen-minimum zones (OMZs), where reductive dissolution and release of Fe from sediments can enrich nearshore bottom waters (Bruland et al., 2005; Elrod et al., 2004; Hong and Kester, 1986; Landing and Bruland, 1987; Luther and Wu, 1997; Moore and Braucher, 2008; Vedamati et al., 2014). A combination of geographical variation in input fluxes such as aeolian deposition, and horizontal mixing at the thermocline, likely drive lateral variability of dissolved Fe (Boyle, 1997; Bruland et al., 1994; Luther and Wu, 1997).

The distinct behavior of dissolved Fe compared to the scavenged endmember or micronutrient metals reflect the specific way in which $Fe^{3+}$ is stabilized in solution in the oceans. Despite its uptake by phytoplankton that contributes to a nutrient-like profile in the upper ocean, inorganic $Fe^{3+}$ is still highly insoluble and increased scavenging and decreasing concentrations with depth should be expected. However, dissolved Fe is maintained at values around 0.6 nM in much of the deep ocean, a phenomenon widely attributed to chelation of $Fe^{3+}$ with organic ligands (Archer and Johnson, 2000; Gledhill and Van den Berg, 1994; Johnson et al., 1997b, 1997a; Kuma et al., 2003, 1996; Millero, 1998; Rue and Bruland, 1995; Van den Berg, 2006; Wu and Luther, 1995). The specific Fe concentration may be controlled by the concentration of strong ligands in the oceans that also occur at levels of 0.4-0.6 nM (Rue and Bruland, 1995; Wu and Luther, 1995). In the model of Johnson et al. (1997a, 1997b), $Fe^{3+}$ abundance up to 0.6 nM is stabilized in solution by chelation with these strong ligands that prevent scavenging, while $Fe^{3+}$ exceeding the concentration of strong ligands is chelated with weaker binding species and thus susceptible to scavenging and precipitation in a similar manner to inorganic $Fe^{3+}$. Some of the geographical variability in dissolved Fe cannot be reconciled however with a simple control by organic ligands (Boyle, 1997; Conway and John, 2014; Fitzsimmons et al., 2015; Luther and Wu, 1997). The 'dissolved' Fe measured in samples passed through 0.45 µm filters contains a component of fine, colloidal material, comprised of authigenic $Fe^{3+}$-oxyhydroxides that is overlooked in endmember ligand-complexation models (Tagliabue et al., 2023). A quasi-equilibrium exists between the dissolved and colloidal phases that is influenced by photochemistry and biota, and varies substantially with depth and geographic location (Cullen et al., 2006; de Baar and de Jong, 2001; Nishioka et al., 2001; Wu et al., 2001). At the same time, the balance of authigenic colloidal versus ligand-bound Fe reflects local balances of biological activity and Fe inputs, and the colloidal pool can represent up to 50 % of all dissolved Fe in some localities (Fitzsimmons et al., 2014a; Kunde et al., 2019). As colloidal Fe is not protected from scavenging in the manner of ligand-bound, dissolved $Fe^{3+}$, part of the geographical variability in Fe scavenging rates, and consequently, variability in the standing stock of Fe in different ocean regions, may reflect the proportion of these two 'dissolved' Fe pools (Moore and Braucher, 2008). In the upper ocean, it has also been recently shown that authigenic Fe can account for over 50 % of the particulate Fe pool (Tagliabue et al., 2023).

The various Fe input fluxes to the oceans are clearly critical to understanding both the standing stock of dissolved Fe, and its spatial distribution. Below we briefly review the sources of Fe to the modern oceans to provide a framework for understanding how variations in these input fluxes might impact the global marine Fe cycle in both the recent and ancient geological past. While we do not discuss in-depth the modern sinks for the dissolved Fe pool in this introductory section, we assume that the total net inputs of Fe to the open ocean are roughly equal to a diffuse flux of Fe-oxides and oxyhydroxides (ferrihydrite, goethite, and magnetite; Feely et al., 1991, Raiswell et al., 2008; von der Heyden et al., 2012) that precipitate from the colloidal Fe pool, settle into sediments across the global seafloor, and undergo diagenetic transformation to other phases during long-term burial. The most striking chemical sedimentary 'sinks' of Fe in the oceans, such as hydrothermal precipitates, and ferruginous and sulfidic sediments formed on highly productive continental margins, should not strictly factor into a discussion of global sinks of modern marine Fe, because they do not derive the bulk of their Fe from the global dissolved Fe pool. They more correctly represent modifiers to local input flux terms; in the case of hydrothermal sediments they are derived directly from Fe-rich fluids, and in the case of anoxic margin sediments they derive their Fe from continental runoff.

*Atmospheric Iron Inputs*

Atmospheric dust deposition is the primary source of Fe to the surface ocean and as such it exerts a primary control on biologically available Fe supply to the photic zone (Fung et al., 2000; Jickells et al., 2005; Mahowald et al., 2005; Moore et al., 2001). Upper ocean dust loading is closely tied to distance from terrestrial desert dust source regions (Jickells et al., 2005; Mahowald et al., 2009, 2005). The major dust source regions are concentrated in North Africa, the Arab Peninsula, Central Asia, China, North America, and South Africa, with the Southern Hemisphere being largely devoid of significant dust activity (Prospero et al., 2002). In a pattern that largely reflects this hemispheric disparity, the highest dust Fe fluxes to the surface oceans, on the order of $10^{-4}$ mol Fe/m$^2$/yr and reaching values as high as $1.4 \times 10^{-2}$ mol Fe/m$^2$/yr, are found in the North and Tropical Atlantic and North Pacific oceans, downwind of major dust source regions, while the Eastern Equatorial Pacific and Southern oceans experience lower modern dust loadings on the order of $10^{-5}$ mol Fe/m$^2$/yr (Fung et al., 2000; Jickells et al., 2005; Mahowald et al., 2009, 2005). As a result, an estimated 8 times more dust enters the Northern than Southern hemispheric ocean regions (Duce and Tindale, 1991). The flux of dust to the oceans also varies temporally, for example in response to seasonal dust plumes (Fitzsimmons et al., 2015), decadal droughts in dust-source regions that can drive four-fold changes in dust flux (Prospero and Lamb, 2003), and up to glacial-interglacial timescales where ice core records show highly (100-fold) variable and elevated dust concentrations during the last glacial maximum (LGM; De Angelis et al., 1987). Higher dust fluxes in glacial climates likely reflect a combination of higher wind speeds, a weaker hydrological cycle, and exposure of continental shelves at lower sea level (Mahowald et al., 2005). Quantifying the global input flux of atmospheric dust to the dissolved ocean Fe pool consists of two parts. First, quantifying the global average flux of dust Fe to the surface ocean and second, quantifying the fraction of dust Fe that is released into solution in seawater. These controls are discussed in Sect. 3.1.

*Hydrothermal Iron Inputs*

Hydrothermal venting at mid-ocean ridges is another highly significant, but quantitatively uncertain flux of Fe to oceans. Hydrothermal circulation (the percolation of seawater through oceanic crust along the mid-ocean ridge system, and its subsequent heating to over 400°C and chemical modification through reaction with host rocks) produces buoyant, reducing, acidic, metal- and sulfide-rich saline fluids that rapidly rise from the seafloor into the overlying water column from hydrothermal vents (German and Seyfried, 2014). The fluids emitted from hydrothermal vents buoyantly rise within the water column, undergoing turbulent mixing and entrainment of cold seawater, until they reach a level of neutral buoyancy, and are dispersed laterally with oceanic currents. Buoyant plumes undergo rapid chemical modification upon their emergence from the seafloor, as cooling and the entrainment of oxygenated seawater drives massive precipitation of Fe-dominated sulfide and oxide particles, with the result that only a fraction of the initial $Fe^{2+}$, which is the most abundant metal species in the hydrothermal fluids, escapes local sedimentation (German and Seyfried, 2014). As such, while it is calculated that the gross flux of hydrothermal fluids with hundreds of micromolar to several millimolar concentrations of Fe from the seafloor is in the range of 1300-10600 Gg/yr, higher than the global riverine flux of Fe (Elderfield and Schultz, 1996), only a fraction of this flux (Saito et al., 2013; Tagliabue et al., 2010) contributes to the wider dissolved Fe pool in the oceans. Uncertainty remains in quantifying this dissolved Fe flux due to the diversity of vent fluid compositions at different mid-ocean ridge sites. For example, dissolved Fe fluxes calculated assuming a fixed $Fe/^{3}He$ ratio, corresponding to the fast-spreading East Pacific Rise (EPR) (Tagliabue et al., 2010), may underestimate the global flux because slow-spreading ridges that represent ~50% of global ridge length have 80 times higher $Fe/^{3}He$ ratios (Saito et al., 2013). Furthermore, diffuse hydrothermal Fe fluxes away from active vent plume sites, which are difficult to detect and thus may be grossly underestimated, represent an additional, poorly quantified, but substantial fraction of the total, stabilized, hydrothermal Fe flux to the oceans (Baker et al., 2016; Lough et al., 2019).

Similarly, current understanding of the geochemical fluxes and fingerprints of hydrothermal venting is heavily skewed towards venting at mid-ocean ridge settings. However, the generally low temperature (<100°C) venting processes, observed at intraplate seamounts such as Kama'euhuakanaloa (formerly Lō'ihi) in the Hawai'ian island chain (Glazer and Rouxel, 2009; Milesi et al., 2023; Rouxel et al., 2018; Wheat et al., 2000) and Mayotte in the Indian Ocean, may also have regional to global marine biogeochemical impacts that have been less extensively studied (Bennett et al., 2011; Jenkins et al., 2020).

Two broad classes of processes control the fate of hydrothermally vented Fe: near-vent or buoyant-plume mineral precipitation processes and long-lived stabilization processes, which are both discussed in Sect. 4.

*Sedimentary Iron Inputs*

Sediments at continental margins represent a major potential source of Fe to the oceans if even a fraction of sedimentary Fe is released to seawater. The delivery of suspended particles with global river discharge is likely more than $19 \times 10^{15}$ g/yr (Peucker-Ehrenbrink, 2009), around 50 times the global dust input to the oceans (Fung et al., 2000; Jickells et al., 2005). If labile Fe release from sediments was comparable to that for the atmospheric dust flux, these fluxes alone would already imply that the sedimentary flux of Fe to the oceans outweighs all other inputs (Jeandel et al., 2011; Lacan and Jeandel, 2005). Iron mobilization from sediments is largely a redox-controlled process, dominated by the reductive dissolution of insoluble $Fe^{3+}$ minerals in the highly reactive Fe ($Fe_{HR}$) sedimentary pool under reducing conditions, when higher potential oxidants ($O_2$, nitrate, $Mn^{4+}$) have been eliminated by oxidation of organic matter (Froelich et al., 1979; Van Cappellen and Wang, 1996). Microbially-mediated dissimilatory $Fe^{3+}$ reduction (DIR) fueled by organic matter (Lovley et al., 1987), in addition to smaller quantities of reduction by $H_2S$ under sufficiently reducing conditions (Dale et al., 2015; Van Cappellen and Wang, 1996), can generate micromolar to millimolar concentrations of dissolved $Fe^{2+}$ in hypoxic to anoxic sediment porewater that diffuses toward the sediment-water interface and enrich bottom waters in Fe (Elrod et al., 2004). Reducing conditions are established where sediments are rich in organic carbon, so most of the sedimentary Fe flux to the oceans originates on the continental margins, particularly in upwelling zones where organic productivity is high (Dale et al., 2015; Elrod et al., 2004; Johnson et al., 1999, 1997a; Moore and Braucher, 2008). Dissolved and particulate Fe is elevated in seawater up to hundreds of kilometers away from continental margins and particularly in OMZs, highlighting the role that Fe dissolution from sediments, the resuspension of sedimentary material, and lateral transport play in marine Fe supply (Bruland et al., 2005; Conway and John, 2014; Homoky et al., 2016, 2012; Hong and Kester, 1986; Johnson et al., 1999, 1997a; Lohan and Bruland, 2008; Moore and Braucher, 2008; Noffke et al., 2012; Tagliabue et al., 2016; Vedamati et al., 2014).

Global iron cycling biogeochemical models tend to produce estimates for the sedimentary flux of roughly equal to a few times the atmospheric Fe input (Moore and Braucher, 2008; Tagliabue et al., 2014). Bottom water $O_2$ (O2BW) plays an important role in determining the Fe flux from sediments (Elrod et al. 2004; Homoky et al., 2012; Severmann et al., 2010). Iron oxidation at higher O2BW acts to drive down dissolved Fe in porewaters, with a sharp 'tipping point' at around 20 µM O2BW below which porewater dissolved Fe sharply increases (Dale et al., 2015; Scholz et al., 2014). Dale et al. (2015) presented a benthic Fe flux estimate that incorporated $C_{org}$ oxidation and O2BW informed by an empirical water-depth dependence of $C_{org}$ oxidation rate (Burdige, 2007) and models of OMZ impingement on the seafloor (Helly and Levin, 2004). They estimated a global dissolved benthic Fe flux to the oceans of ~8400 ± 4200 Gg/yr, with 6000 ± 3000 Gg/yr, from continental margin sediments (of which shelf and slope setting account for 4000 and 2000 Gg/yr, respectively) and 2300 ±1200 Gg/yr from deep sea (>2000 m water depth) sediments. The major source of uncertainty in these flux estimates is the $Fe_{HR}$ fraction of the sedimentary flux, which can be highly variable (Aller et al., 1986; Dale et al., 2015; Poulton and Raiswell, 2002; Raiswell, 2011). While highlighting that sedimentary Fe sources are perhaps the dominant Fe flux to the deep ocean, Dale et al. (2015) observed that most of this Fe does not reach the photic zone and thus deep marine Fe scavenging processes are likely to be heavily underestimated, with the possible existence of a rapid removal mechanism taking place in near-seafloor, particle-rich layers. Reductive-dissolution-focused benthic Fe supply calculations may yet underestimate global fluxes because they cannot accurately account for non-reductive dissolution (NRD) sources of Fe from more oxidizing sedimentary settings (Homoky et al., 2016, 2013; Klar et al., 2017; Radic et al., 2011).

*Other Iron Inputs*

Most other sources of Fe to the oceans are not thought to be of global significance, but rivers (Bergquist and Boyle, 2006; Escoube et al., 2015, 2009; Fantle and DePaolo, 2004), glaciers (Bhatia et al., 2013; Hawkings et al., 2014; Raiswell et al., 2006; Zhang et al., 2015), and coastal groundwater discharges (Rouxel et al., 2008b; Rouxel and Auro, 2010; Roy et al., 2012, 2010; Windom et al., 2006) represent point sources that may locally affect marine Fe concentrations and isotopic signatures. Rivers and glaciers ultimately represent sources of Fe to the continental margins where the processes described above in the context of the sedimentary Fe input will act to remobilize Fe, up to hundreds of km away from the coastline (Johnson et al., 1997a). Treating such fluxes as a separate Fe input to the global oceans might result in double counting of fluxes described above.

Iron delivery from infalling extraterrestrial dust has also been proposed as a non-negligible contributor of Fe to the oceans (Johnson, 2001; Reiners and Turchyn, 2018; Rudraswami et al., 2021). As extraterrestrial dust falls through Earth's atmosphere, it undergoes friction heating in the upper atmosphere that causes evaporation of ~90 % of the mass of the infalling particles (Taylor et al., 1998). The evaporated material reaccretes as particles of meteoritic 'smoke' with diameters on the order of 1 nm that are likely highly soluble and thus bioavailable (Hunten et al., 1980; Johnson, 2001). Unlike the point sources detailed in the paragraph above, the extraterrestrial dust supply should be fairly spatially uniform. Assuming a widely accepted global extraterrestrial flux of $4 \times 10^7$ kg/yr, an average meteoritic [Fe] of 25 wt %, bioavailability of the 90 % of the original dust that evaporated, and uniform distribution over Earth's surface, which is 70 % ocean, the total bioavailable extraterrestrial Fe flux to the modern oceans should be approximately 6 Gg/yr (Reiners and Turchyn, 2018), much lower than the global aeolian Fe input, depending on solubility assumptions. If deposited uniformly, this flux might be non-negligible in aeolian-dust-starved surface regions such as the Southern Ocean (Johnson, 2001).

*Windows to the ancient Earth from iron cycling in anoxic ocean and lake environments*

Environments exist on the modern Earth that offer a glimpse into the complexities of Fe cycling in less oxidizing conditions ranging from suboxic to highly euxinic. Among these sites are the Black Sea, Cariaco Trench, Peru upwelling region (Fig. 5), and numerous redox-stratified lakes (Fig. 6). Below, we briefly review the state of knowledge of Fe cycling in some of these modern analogs for ancient marine environments, to provide a foundation for speculating on water column redox and geochemical structure, and dominant geochemical processes, that affected the ocean DIB cycle prior to the Phanerozoic.

The Black Sea (Fig. 5A) is the largest anoxic basin on Earth (Caspers, 1957) and it has become a foundation of the current understanding of marine redox geochemistry in anoxic oceans in deep time (Lyons et al., 2009; Lyons and Severmann, 2006). With waters supplied to the surface from terrestrial runoff and an input of salty Mediterranean Sea water via the Dardanelles and Bosphorus at depth, anoxia in the basin is maintained by a strong density stratification that isolates deep waters from the oxygenated atmosphere. The density stratification is driven by a contrast in surface (17.2-18.3 ‰) and deep water (22.4 ‰) salinity that prevents downward mixing of cold surface waters in winter (Spencer and Brewer, 1971). The redox boundary resulting from this physical stratification is at around 110 m in the central basin, below which dissolved $O_2$ concentrations fall to zero and free $H_2S$ appears, and increases with depth to ~400 μM (Rolison et al., 2018), as a product of anoxic microbial sulfate reduction at the expense of sinking organic carbon. The deep Black Sea basin is therefore an euxinic environment, where $H_2S$ is the dominant redox species with which neither dissolved $O_2$ nor $Fe^{2+}$ can coexist.

A substantial portion of particulate Fe in the Black Sea is lithogenic silicate material (Muramoto et al., 1991; Spencer et al., 1972) that is not highly reactive, while the dissolved pool of Fe in the Black Sea also has a sedimentary origin, ultimately being derived from remobilization of highly reactive Fe in sediments on the oxic shelf of the basin (Lyons and Severmann, 2006; Severmann et al., 2008; Wijsman et al., 2001; Lenstra et al. 2019). Both reactive Fe speciation (Lyons and Severmann, 2006; Wijsman et al., 2001) and Fe isotopic (Severmann et al., 2008) studies of shelf and basinal sediments indicate that reductive dissolution of $Fe^{3+}$ oxyhydroxides in shelf sediments occurs during diagenesis and mobilizes $Fe^{2+}$ from the shelf. Transport along isopycnal surfaces around the oxic-anoxic interface then shuttle the liberated $Fe^{2+}$ that survives partial reoxidation to the open basin (Lyons and Severmann, 2006; Severmann et al., 2008).

The redox structure of the Black Sea creates a distinct dissolved Fe profile in the water column. In the oxygenated, upper part of the water column, Fe concentrations are on the order of 1 nM, although these can be elevated to a few nM in nearshore and surface waters (Lewis and Landing, 1991; Rolison et al., 2018). Iron exists as 'free' $Fe^{3+}$ along with some colloidal and complexed ferric phases in the oxic portion of the water column. Starting at intermediate depths, corresponding to the suboxic zone a few tens of meters above the appearance of $H_2S$, dissolved Fe begins to increase with depth, reaching ~20 nM at the onset of the sulfide interval, and reaching peak values of 440 nM at depths of 40-50 m below the top of the sulfide interval (Lewis and Landing, 1991; Rolison et al., 2018). Iron concentrations drop sharply over a few hundred meters beneath this concentration peak.

The two processes predominantly responsible for the dissolved Fe concentration profile are redox cycling of Fe around the depth of the oxic-anoxic interface, and the precipitation and removal of Fe sulfides at depth below the Fe concentration peak where these phases are supersaturated in solution (Calvert and Karlin, 1991; Canfield et al., 1996; Konovalov et al., 2004; Lewis and Landing, 1991; Lyons and Berner, 1992; Lyons and Severmann, 2006; Muramoto et al., 1991; Raiswell and Berner, 1985; Rolison et al., 2018; Severmann et al., 2008; Wijsman et al., 2001). In the water column, upwardly advected $Fe^{2+}$ is rapidly oxidized into $Fe^{3+}$ oxyhydroxides, via reaction with $O_2$ or sinking $Mn^{4+}$ oxides that precipitate at shallower depths (Konovalov et al., 2004; Lewis and Landing, 1991; Rolison et al., 2018; Spencer et al., 1972; Spencer and Brewer, 1971). Upon sinking below the oxic-anoxic interface, into the sulfidic zone of the water column, reductive dissolution of $Fe^{3+}$ oxyhydroxides by sulfide takes place, producing the sharp peak in dissolved Fe concentrations that is partitioned between free $Fe^{2+}$ and colloidal Fe-S (Lewis and Landing, 1991). Here, Fe and $H_2S$ concentrations far exceed saturation levels for Fe-sulfides such as mackinawite, greigite, and pyrite, with observations of pyrite framboids in sediment traps and isotopic systematics both indicating that pyrite is the dominant sulfide phase formed (Konovalov et al., 2004; Lewis and Landing, 1991; Muramoto et al., 1991; Rolison et al., 2018). The degree of pyritization (percentage of total reactive Fe in the form of pyrite) in Black Sea euxinic basin sediments indicates that the majority of reactive Fe deposited is precipitated in the water column, as opposed to diagenetic transformation of other reactive Fe species, and that the formation of pyrite in the Black Sea water column is Fe-limited (Calvert and Karlin, 1991; Canfield et al., 1996;

Lyons and Berner, 1992; Raiswell and Berner, 1985).

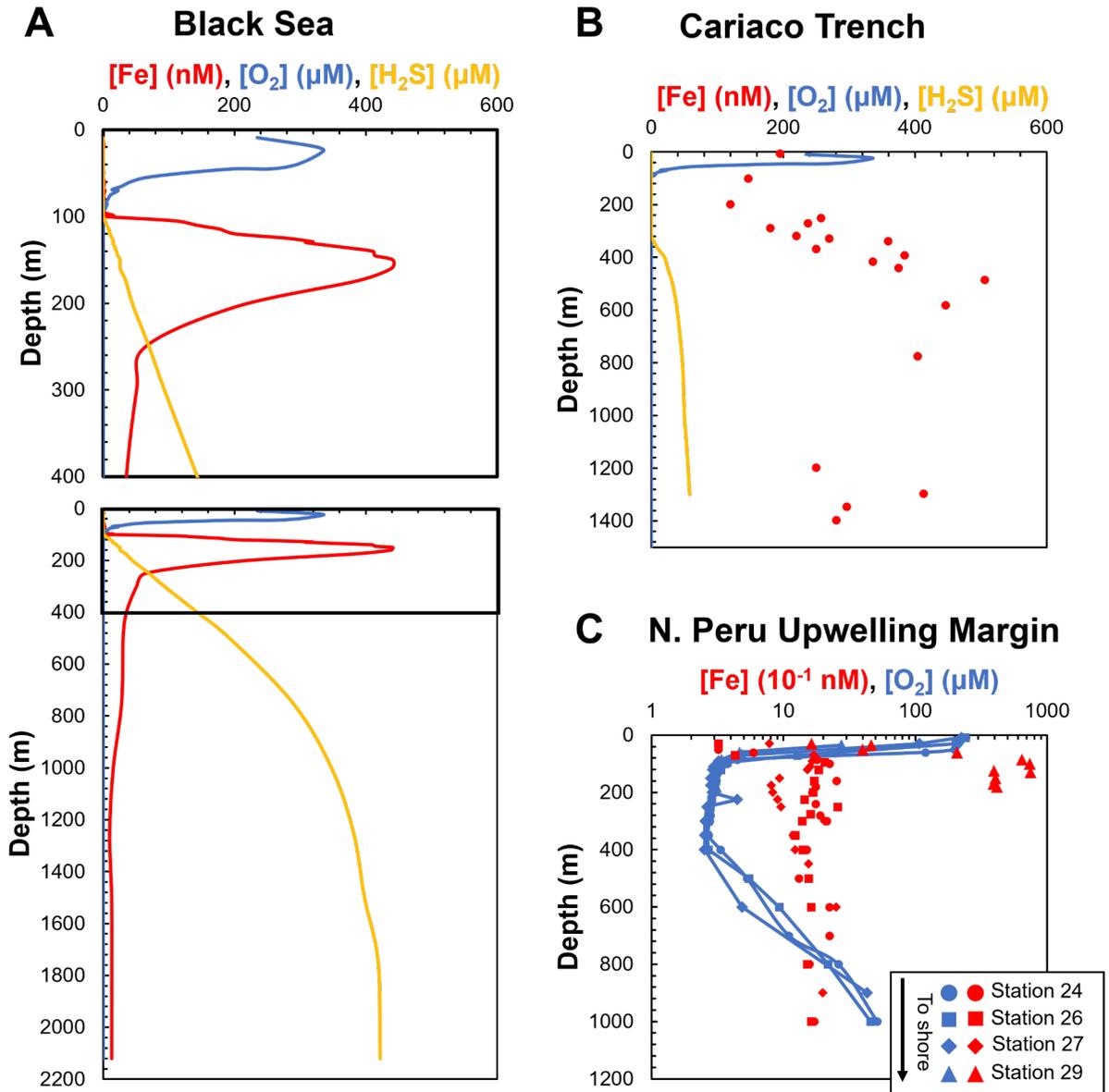

**Fig. 5.** Redox species ([Fe], [$O_2$], and [$H_2S$]) profiles in A: The Black Sea, data from Rolison et al. (2018). B: The Eastern Cariaco Trench, data from Bacon et al. (1980) and Zhang and Miller (1993). C: The North Peru Upwelling Margin, data from Vedamati et al. (2014).

The Cariaco Trench (Fig. 5B) is a ~1400 m depth depression in the continental shelf off Venezuela, separated from the Caribbean Sea to the northeast by a sill less than 150 m deep that includes Islas Tortuga, Margarita, Cubagua, Coche, and Farallon Centinela. The sill encloses a 100-mile-long, 40-mile-wide trench and prevents free exchange of waters deeper than 150 m (Richards, 1975; Richards and Vaccaro, 1956). Waters below that depth are devoid of $O_2$ and $H_2S$-rich (Richards, 1975, 1960; Richards and Vaccaro, 1956; Zhang and Millero, 1993). A more gradual density gradient of the Cariaco Trench in comparison to the Black Sea means that the redox transition depth shows substantial temporal variability (Percy et al., 2008; Richards, 1975;

Scranton et al., 2001, 1987; Zhang and Millero, 1993). The redox structure of the Cariaco Trench broadly resembles that of the Black Sea (Bacon et al., 1980; Lewis and Landing, 1991; Spencer et al., 1972; Spencer and Brewer, 1971). Dissolved $O_2$ decreases with depth to an oxic-anoxic transition zone marked by the appearance of $H_2S$, at depths of 220-375 m, depending on sampling date and location (Bacon et al., 1980; Fanning and Pilson, 1972; Percy et al., 2008; Richards and Vaccaro, 1956; Zhang and Millero, 1993). At the top of this transition, a suboxic zone defined by $O_2$ and $H_2S$ both being < 1 μM is observed with a thickness ranging from 5 to 50 m (Percy et al., 2008). Beneath the transition to anoxic conditions, dissolved Fe concentrations increase sharply to 300-600 nM, as solid $Fe^{3+}$ oxyhydroxides are reduced and solubilized by reaction with $H_2S$ and dissimilatory reactions with organic carbon (Bacon et al., 1980; Jacobs et al., 1987; Percy et al., 2008; Zhang and Millero, 1993). A slightly shallower peak in dissolved Mn concentrations, overlain by a particulate $Mn^{4+}$ oxide peak is observed, likely reflecting a similar interaction between the oxidation of upwelled Mn and subsequent redissolution upon sinking as is seen in the Black Sea (Bacon et al., 1980; Lewis and Landing, 1991). Dissolved $Fe^{2+}$ concentrations fall slightly with depth in the anoxic layer, but remain in the 100s of nM level (Bacon et al., 1980; Jacobs et al., 1987; Percy et al., 2008), in contrast to the very sharp decline in concentration seen in the Black Sea (Konovalov et al., 2004; Lewis and Landing, 1991; Muramoto et al., 1991; Rolison et al., 2018). Both thermodynamic calculations (Jacobs et al., 1987; Percy et al., 2008; Scranton et al., 2001) and studies of the laminated sediments at the basin floor (Lyons et al., 2003; Yarincik et al., 2000) indicate that Fe sulfides precipitate in the euxinic deep waters of the Cariaco Trench. However, the maximum $H_2S$ concentration reached at depth in the basin, controlled by a combination of deep-water sulfate reduction and diffusion from sediments into bottom water (Fanning and Pilson, 1972; Scranton et al., 1987), is an order of magnitude lower than in the Black Sea, such that sulfide mineral saturation is only reached at higher dissolved Fe levels. Buffering of the Cariaco Trench Fe profile to saturation with respect to sulfide minerals is further underlined by the relatively rapid readjustment of water column chemistry back to a near-constant ion activity product in the aftermath of climatically or tectonically influenced changes in the terrigenous Fe flux to the basin (Percy et al., 2008; Scranton et al., 2001).

Significant changes in the depth of the oxic-anoxic transition zone and deep-water $H_2S$ concentration in the Cariaco Trench have taken place on years-to-decadal timescales and reveal the basin to be in a non-steady state. Hydrogen sulfide concentrations in the range of 40-75 μM were observed since the 1970s, with upward shifts of the oxic-anoxic transition zone accompanying increased $H_2S$, and vice-versa (Holmén and Rooth, 1990; Percy et al., 2008; Richards, 1975; Scranton et al., 2001, 1987; Zhang and Millero, 1993). Likely drivers include changes in ventilation by cold or hypersaline waters originating on the shelf or sill (Fanning and Pilson, 1972; Holmén and Rooth, 1990; Richards, 1960; Scranton et al., 1987), input from or disturbance of $Fe^{3+}$ oxyhydroxide-rich sediments that enhanced $H_2S$ oxidation and sulfide mineral precipitation (Jacobs et al., 1987; Percy et al., 2008; Scranton et al., 2001; Zhang and Millero, 1993), and seasonal or longer term changes in upwelling-driven productivity. Evidence for a ventilation-driven influence on deep-water $H_2S$ includes correlations with conservative parameters like salinity and potential temperature, the presence of tritium in deep waters, and the appearance of non-zero $O_2$ levels within the anoxic transition zone (Fanning and Pilson, 1972; Holmén and Rooth, 1990; Richards, 1975, 1960; Scranton et al., 1987; Zhang and Millero, 1993). Meanwhile, changes in $H_2S$ unaccompanied by measurable changes in hydrographic parameters have been linked to Fe loading in the aftermath of earthquakes and inferred suspension of terrigenous sediments in turbidity flows (Percy et al., 2008; Scranton et al., 2001; Zhang and Millero, 1993).

Perhaps unsurprisingly in light of this short-term variability, various estimates of water residence time in the basin return consistently small values, on the order of 100 years (Deuser, 1973; Holmén and Rooth, 1990; Richards and Vaccaro, 1956; Scranton et al., 1987; Zhang and Millero, 1993).

The Peru upwelling region (Fig. 5C) lies within an extensive oxygen-minimum zone (OMZ) in the eastern Southern Pacific Ocean, representing an example of Fe cycling in a reducing marine environment fully connected to the open ocean. The OMZ is a consequence of intense oxygen consumption resulting from high surface organic matter production and degradation, maintained by strong wind-driven upwelling of poorly oxygenated, nutrient-rich Pacific deep-waters, and sluggish ventilation from surface waters (Brink et al., 1983, 1980; Dengler, 1985; Fuenzalida et al., 2009; Gunther, 1936; Kessler, 2006; Wyrtki, 1962; Zuta and Guillén, 1970). The OMZ, defined by waters with dissolved $O_2$ < 20 μM, encompasses roughly <100 m to 700 m water depth, and extends up to ~1000 km westward from the shoreline of Peru between 5 and 15°S (Bruland et al., 2005; Fuenzalida et al., 2009; Hong and Kester, 1986). Dissolved $O_2$ in the core of the OMZ falls below 2 μM (Fuenzalida et al., 2009; Vedamati et al., 2014), and where it impinges on the continental shelf and slope to water depth as shallow as 50 m, a fine-grained mud lens is deposited with up to 10 wt% organic matter (Reimers and Suess, 1983; Scheidegger and Krissek, 1983).

Waters within the OMZ on the Peru margin are extremely Fe-rich. Waters overlying the continental shelf typically contain 20-50 nM of dissolved Fe, predominantly as $Fe^{2+}$, which decreases offshore to a few nM just past the shelf break 200 km westward, and to near-global ocean values of < 1 nM by 400 km from the shore (Bruland et al., 2005; Hong and Kester, 1986; Vedamati et al., 2014). Within the OMZ, Fe is retained in solution by reducing conditions that stabilize soluble $Fe^{2+}$. Water column transects show that Fe originates from underlying anoxic sediments originally delivered to the shelf by continental runoff, as near-bottom water dissolved Fe concentrations can exceed 100 nM, while little Fe enrichment is seen in waters off Southern Peru, where the shelf is only 10 km wide and provides an inadequate source of sedimentary material (Bruland et al., 2005; Vedamati et al., 2014). Furthermore, a co-enrichment of $Fe^{2+}$ with nitrite within the nearshore OMZ further suggests a sedimentary origin for dissolved $Fe^{2+}$ in the water column related to organic matter decomposition (Hong and Kester, 1986; Vedamati et al., 2014). Sediment profiles beneath the Peru upwelling highlight the role of diagenesis in delivery of Fe to the overlying water column. Within sediments underlying the permanent OMZ, pore water $Fe^{2+}$ concentrations reach maxima at or within a few cm of the sediment-water interface with typical values of 20-30 nM (and up to 80 nM), reflecting anaerobic organic matter degradation coupled to reduction of $Fe^{3+}$ oxyhydroxides delivered to shelf sediments (Noffke et al., 2012; Scholz et al., 2014, 2011; Suits and Arthur, 2000). Due to the abundance of reactive Fe available for respiration, sulfate reduction occurs deeper in the sediment column, and free $H_2S$ generated in porewaters cannot extend above the sediment-water interface because it is removed via Fe sulfide precipitation, resulting in a ferruginous redox state of the overlying water column (Noffke et al., 2012; Suits and Arthur, 2000). Total iron to aluminum ($Fe_T$/Al) ratios of sediments within the OMZ are lower than the local crustal (andesitic) value representative of the continental runoff to the shelf, which confirms that reduction of $Fe^{3+}$ oxyhydroxides in the sediment during organic matter degradation drives a net removal of Fe from sediments to the overlying water column (Noffke et al., 2012; Scholz et al., 2014, 2011). Conversely, in a slope setting further westward, at water depths below the lower boundary of the OMZ, $Fe_T$/Al ratios of sediments are higher than the crustal value, because after $Fe^{2+}$-rich waters leave the core of the permanent OMZ and encounter more oxygenated conditions, $Fe^{2+}$ is effectively oxidized to insoluble $Fe^{3+}$ oxyhydroxides that are deposited in sediments and enrich the reactive Fe-pool (Noffke et al., 2012; Scholz et al., 2014, 2011). Iron isotopic data indicate that this 'open ocean iron shuttle' that transports Fe from the shelf to the slope is a 'leaky' process with only partial re-oxidation of shuttled $Fe^{2+}$ taking place (Scholz et al., 2014).

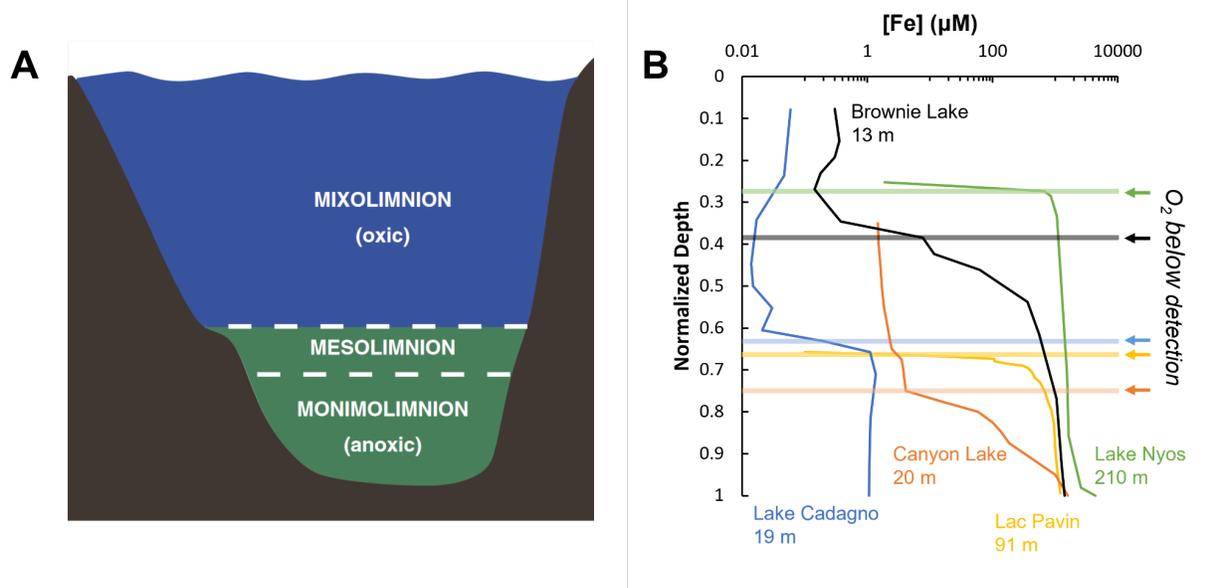

**Fig. 6.** Schematic and chemical profiles of meromictic lakes with anoxic depths. A: Water column schematic of Lac Pavin, after Busigny et al. (2014), illustrating the principal layers in a meromictic lake. B: Dissolved [Fe] profiles of a selection of meromictic lakes with anoxic deep waters. Shaded horizontal lines show the base of the oxycline for each lake, beneath which [$O_2$] is below detection. Data from Busigny et al. (2014) (Lac Pavin), Ellwood et al. (2019) (Lake Cadagno), Teutsch et al. (2009) (Lake Nyos), and Lambrecht et al. (2018) (Brownie and Canyon Lakes).

Modern lakes with redox-stratified water columns (Fig. 6) have been extensively studied as analogue sites for past anoxic oceans despite some of the features of lakes that make their geochemistry unique from the marine realm (Degens and Stoffers, 1976). One key distinction between the modern oceans and lakes is that in many cases the latter do not have high sulfate concentrations, so that when they become anoxic at depth, $Fe^{2+}$ is able to accumulate as the dominant electron donor species (rather than $H_2S$). Widespread ferruginous environments are not established in sulfate-rich modern oceans, but may have dominated ancient oceans for much of Earth history, making $Fe^{2+}$-rich stratified lakes a valuable window into the past, despite their numerous uniquely non-marine features (Swanner et al. 2020).

Lakes that are permanently stratified, or 'meromictic', offer the best, most stable analogue environments for study. Meromictic lakes are defined by their separation into an upper mixed layer (mixolimnion), and a lower layer (monimolimnion) that has a higher density and is resistant to mixing with the oxygenated atmosphere above. Just a few hundred meromictic lakes are thought to exist worldwide, with only a few dozen documented examples of naturally occurring, circumneutral-pH lakes of ferruginous character (Swanner et al., 2020). No two lakes are the same, and their chemistry in many cases reflects composition of bedrock lithologies, with, for example, karstic lakes having high dissolved inorganic carbon loads (Camacho et al., 2017; Lambrecht et al., 2018; Swanner et al., 2020).

The general redox structure of ferruginous meromictic lakes features an oxygenated, $Fe^{2+}$-poor mixolimnion in communication with the oxygenated atmosphere and often inhabited by cyanobacteria, and an $Fe^{2+}$-rich, anoxic monimolimnion. A well-defined boundary or redoxcline often separates these two layers, where reaction of $O_2$ with $Fe^{2+}$ to form insoluble $Fe^{3+}$-oxyhydroxides drives the concentration of both species down close to zero (Busigny et al., 2014; Crowe et al., 2008; Davison, 1993; Emmenegger et al., 1998; Lambrecht et al., 2018; Swanner et al., 2020; Teutsch et al., 2009; Viollier et al., 1995). Universally, $Fe^{2+}$ concentration in the monimolimnion of ferruginous lakes exceeds modern marine concentrations with values ranging from approximately 100 μM to several 1000 μM in different localities (Swanner et al., 2020). The $Fe^{2+}$-$O_2$ profiles of ferruginous lakes can be simply understood in terms of reactions involving the oxidation and reduction of Fe. Removal of $Fe^{2+}$ above a certain depth reflects oxidation by reaction with $O_2$ (Emmenegger et al., 1998), and the depth at which $Fe^{2+}$ approaches near zero concentrations can be estimated with simple reaction-transport models (Heard et al., 2020; Teutsch et al., 2009). Below the redoxcline, sinking $Fe^{3+}$-oxyhydroxides undergo reduction driven by microbial DIR coupled to sinking organic matter (Davison, 1993). Internal redox cycling in ferruginous lakes can be highly efficient, with cases of up to 90% recycling of Fe between the redoxcline and monimolimnion observed (Campbell and Torgersen, 1980). However, this process is not a perfectly closed loop, and thus external Fe supplies are balanced by the long-term burial of Fe in sediments (Davison, 1993; Nürnberg and Dillon, 1993). In some lakes like Cadagno (Switzerland) a very fast turnover of reduction and oxidation processes is seen (Berg et al. 2016).

Studies of iron-bearing minerals in ferruginous lakes can provide a window into the diversity of authigenic and diagenetic mineral assemblages that may have formed in ferruginous early oceans. Mineral formation in the water column of ferruginous lakes is broadly controlled by the precipitation of ferric oxides and oxyhydroxides at and above the redoxcline following oxidation by $O_2$ and potentially anaerobic phototrophs (Emmenegger et al., 1998; Lambrecht et al., 2018; Swanner et al., 2020; Viollier et al., 1995), and precipitation of various $Fe^{2+}$-dominated phases in the reducing monimolimnion (Cosmidis et al., 2014; Havig et al., 2015; Lambrecht et al., 2018; Michard et al., 1994; Swanner et al., 2020; Viollier et al., 1995). The nature of the minerals formed in the monimolimnion of lakes depends on water column chemistry (Lambrecht et al., 2018). Siderite ($FeCO_3$) is a ubiquitous precipitate owing to the prevalence of DIC either from leaching of carbonate bedrock lithologies or remineralization of organic matter, and should be a robust indicator of past ferruginous conditions in limnological records (Swanner et al., 2020). Phosphate-rich, ferruginous water columns are both predicted (Lambrecht et al., 2018; Michard et al., 1994), and observed (Cosmidis et al., 2014) to precipitate the ferrous Fe-phosphate phase, vivianite - $Fe_3(PO_4)_2 \cdot 8(H_2O)$, in anoxic deep waters, with mixed valence Fe-phosphates also forming closer to the redoxcline. In meromictic lakes with appreciable natural (*e.g.*, Green Lake, New York) or anthropogenic (*e.g.*, Brownie Lake, Minnesota) sulfate, bottom waters can become seasonally or permanently euxinic, and Fe sulfides, predominantly pyrite, can form (Busigny et al., 2014; Havig et al., 2015; Herndon et al., 2018; Lambrecht et al., 2018; Viollier et al., 1995).

A key research area in using ferruginous lakes in early Earth analogue studies has been to test the viability of photoferrotrophy (anoxygenic photosynthesis that uses $Fe^{2+}$ as an electron donor) as a driver of primary productivity and widespread chemical deposition of Fe-bearing minerals before the rise of oxygen. The existence of a photoferrotrophic metabolism and its potential role in early biogeochemical cycling have been known and discussed for a few decades (Kappler and Newman, 2004; Konhauser et al., 2002; Widdel et al., 1993). Photoferrotrophy has been sought predominantly in lakes where the depth of illumination extends below the depth of the redoxcline, such that upwelling anoxic and $Fe^{2+}$-rich waters can receive a significant photon flux before encountering $O_2$. In general, smaller, temperate lakes with deep redoxclines, high productivity, or high turbidity due to, for example, humus loading, are not good targets for seeking out photoferrotrophy (Lambrecht et al., 2018; Swanner et al., 2020). Lakes in which photoferrotrophy has been proposed on the basis of 16S RNA gene sequencing that revealed the presence of anoxygenic photosynthesizers such as green sulfur bacteria (GSB) and purple sulfur bacteria (PSB) within the anoxic, $Fe^{2+}$-rich monimolimnion include Lake Matano, Indonesia (Crowe et al., 2008); Lake La Cruz, Spain (Camacho et al., 2017; Walter et al., 2014), and Lake Cadagno, Switzerland (Berg et al., 2016). However, as noted by Swanner et al. (2020), demonstrating active photoferrotrophy requires the co-occurrence of these genes with evidence for $Fe^{2+}$- and light-dependent C-fixation, because other electron donors like $H_2S$ and $H_2$ are also utilized by anoxygenic photosynthesizers. For example, in Lake Matano, Crowe et al. (2014) reported that low-level C-fixation occurring in deep, ferruginous waters was attributable to sulfur oxidation even at the micromolar sulfide level encountered in the water column, despite the abundance of $Fe^{2+}$. In Lake La Cruz, Walter et al. (2014) did identify $Fe^{2+}$- and light-dependent C-fixation in a culture sampled from a deep, second chlorophyll peak beneath the redoxcline. A corresponding small particulate $Fe^{3+}$ peak was observed at this level, and the occurrence of 'anoxic $Fe^{3+}$ peaks' as an indicator of photoferrotrophy can be entertained, but the role of cyanobacteria even at these deeper levels remains equivocal, as does the significance of photoferrotrophy in ferruginous lake biogeochemistry in general (Swanner et al., 2020 and references therein). Cryptic redox cycling may also mask the efficiency of photoferrotrophy if the re-reduction of $Fe^{3+}$ occurs faster than $Fe^{2+}$ oxidation (Kappler et al. 2021).

## 3. Aeolian and riverine export from the continents

In modern oxic environments and at near-neutral pH, $Fe^{2+}$ released by alteration of crustal rocks is rapidly oxidized into $Fe^{3+}$, which polymerizes to form insoluble $Fe^{3+}$ oxyhydroxides such as ferrihydrite. This explains why ferruginous waters are extremely rare in modern terrestrial surface environments, occurring only in low-pH systems such as acid mine drainage where $Fe^{2+}$ oxidation kinetics are slow. Due to its low solubility, the residence time of iron in the modern oceans is short and its distribution in the oceans is highly heterogeneous. Iron is used by phytoplankton for enzyme catalysis and electron transfer, and it is a limiting nutrient in large swaths of the world's oceans. Primary productivity in those regions (Fig. 3) depends on iron delivery through riverine and aeolian transport (Sect. 2). Export of dissolved iron from the continents to the oceans is complex. It involves chelation (and stabilization) of dissolved iron by organic molecules, and solubilization of iron from aeolian dust by complex solid-water chemistry in the atmosphere. Despite our limited understanding of past conditions in Earth's atmosphere and oceans, reconstructing the continental iron export flux under reducing conditions might be easier because the behavior of dissolved $Fe^{2+}$ is more straightforward and it can be more easily scaled to the flux of other elements such as $Mn^{2+}$ and $Mg^{2+}$. We start below by examining quantitatively continental export of iron at present and then proceed to estimate plausible fluxes for the Archean and Proterozoic.

### 3.1. Aeolian transport

While rivers deliver large amounts of iron to the oceans, much of that is captured in coastal regions through breakup of chelated complexes in marine waters, oxidation, and biological consumption. Aeolian transport plays a more significant role in exporting iron from continents to the open ocean over distances of thousands of kilometers (Fig. 7; Hamilton et al., 2020; Ito et al., 2019; Jickells et al., 2005; Myriokefalitakis et al., 2018; Scanza et al., 2018). Aeolian dust is primarily produced in arid regions that lack vegetation and have poor soil cohesion. Iron-bearing aerosols are also produced by fires and are released by human activity such as fossil fuel burning. The dust particles that make it to the open oceans are small, mostly less than 10-20 µm in size. Some dust particles can interact with acidic water during transport through the atmosphere, releasing biogeochemically available iron that would otherwise be locked in non-reactive minerals. Photochemistry and organic compounds can also play some roles in iron solubilization by enabling $Fe^{3+}$ reduction. Calculation of the amount of dissolved iron delivered to the oceans by aeolian transport is nontrivial as it involves multiple uncertain steps (dust generation, aeolian transport, mineral-water interactions, dust deposition, and release of biogeochemically available iron). Tagliabue *et al*. (2016) compared aeolian transport of iron in 13 global ocean iron biogeochemistry models. The flux of dissolved iron from aeolian dust in these models ranges between 80 and 1800 Gg yr$^{-1}$, with arithmetic and geometric means of 670 and 340 Gg yr$^{-1}$, respectively. We are interested in evaluating how the flux of iron to the oceans could have been different in the past, so it is important for this purpose to understand the parameters that control this flux and the origin of the large scatter between different estimates. The global dust emission rate for dust particles smaller than 10 or 20 µm in size is estimated to be 1600 to 3500 Tg yr$^{-1}$, of which only 197 to 686 Tg yr$^{-1}$ (wet and dry deposition combined; mean ~440 Tg yr$^{-1}$) is deposited in the oceans (Wu et al., 2020). The Fe concentration in atmospheric dust (equivalent to that in the upper continental crust, which is its main source) is ~3.5 wt% (Taylor and McLennan, 1985) and laboratory experiments show that only ~1-10 % of iron can be dissolved (Aumont et al., 2003; Fung et al., 2000; Jickells and Spokes, 2001). The present aeolian flux of labile iron to the oceans is therefore ~460 Gg yr$^{-1}$, with a large uncertainty of ~100 to 2000 Gg yr$^{-1}$. This back-of-the-envelope calculation shows that the two most uncertain parameters in this calculation are the rate of dust deposition to the oceans and the fraction of soluble iron. We discuss below the different controls on Fe delivery in the Archean and Proterozoic, including paleogeography, the exposure of continents above sea level, and redox controls on Fe mobility; bearing in mind that while uncertainties are significant it is less of an issue than in the modern day since other sources of iron presumably played a much more significant role in the DIB cycle.

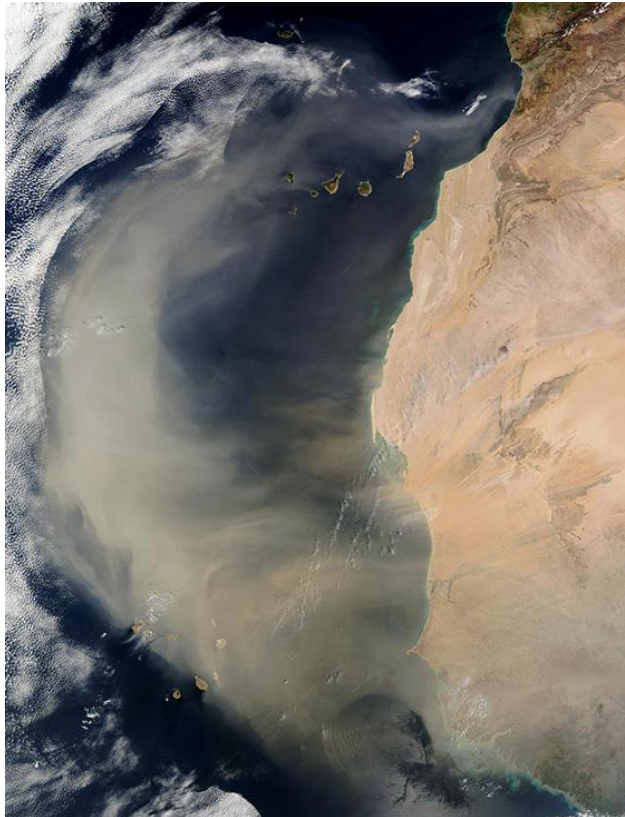

**Fig. 7.** Dust storm moving off of Africa into the North Atlantic on 02 March 2003 captured by the Moderate Resolution Imaging Spectroradiometer (MODIS) instrument, aboard NASA's Terra and Aqua satellite (Credit: NASA MODIS Rapid Response Team).

Earth has seen several cycles of supercontinent buildup and breakup over the past ~3 Gyr, which going backwards from present are named Pangea, Gondwana, Pannotia, Rodinia, Columbia, and Kenorland (Meert, 2012; Pesonen et al., 2003; Scotese, 2009; Zhao et al., 2006; Zhao et al., 2018). Clear peaks in the age distribution of detrital zircons corresponding to the supercontinent cycle give credence to the existence of the oldest supercontinents (supercratons), for which evidence is scant (Hawkesworth et al., 2010). Earth's paleogeography changed significantly during the Archean and Proterozoic, most likely affecting aeolian dust transport via changes in wind, temperature, and precipitation patterns. For example, the argument has been made that the Superia supercraton might have been covered by large epicontinental seas, such that the hydrological cycle was different, involving shorter-distance atmospheric water transport as landmasses were smaller (Bindeman et al., 2018; Spencer et al., 2019). This would have correspondingly reduced exposure of erodible material to wind transport, but increased the chance that wind-blown dust would end up in the sea. Continental emergence at the end of the Archean (Flament et al., 2013; Korenaga et al., 2017; Rey and Coltice, 2008) likely also affected dust export to the oceans. Colonization of emerging landmasses by plants during the Phanerozoic had a stabilizing effect on soils and likely affected aeolian dust transport as well.

The atmosphere during stage #3 (between GOE and NOE) contained some oxygen, but at a lower level than what it is at present, with $pO_2$ values between 0.001 and 0.1 PAL proposed (Sect. 1). In the Archean (stage #1) atmosphere, $pO_2$ was extremely low, and methane could have been relatively abundant with concentrations of 5000 ppm suggested (Catling and Zahnle, 2020).

The reductive power of the atmosphere in the Archean would have promoted iron mobilization. The lower luminosity of the Sun calls for higher $p$CO$_2$ in the Archean, possibly on the order of ~0.1 bar (~300 PAL) (Catling and Zahnle, 2020), which would have acidified rainwater to a pH of ~5 (Hao et al., 2017), thereby promoting leaching of iron and other nutrients from minerals undergoing subaerial weathering. The pH of modern rainwater is also lowered by anthropogenic NOx and SO$_2$, so this CO$_2$-induced acidification most likely had a modest effect. More significant is the reducing state of the atmosphere in the Archean that would have enabled mobilization of Fe$^{2+}$ leached from minerals, which is limited at present by rapid oxidation. It is difficult to assess how a reducing atmosphere could have affected iron dissolution in aerosols as the redox state of iron, organic complexation, and photochemistry would have been different. As a first pass on this question, we calculate the fraction of dissolvable iron under a reducing atmosphere by assuming that all Fe$^{3+}$ in modern aerosols represents highly reactive secondary Fe that would have been leachable. The average Fe$^{3+}$/Fe$_{tot}$ ratio of modern ambient aerosols is ~0.6 (geometric mean from the compilation of Mao et al. 2017). This oxidized iron is difficult to mobilize, but most of it went through some redox cycling by water-rock interactions, meaning that it might have been labile in the Archean and Proterozoic. Approximately ~3% of total iron in aeolian dust can be dissolved at present (Journet et al., 2008; Marcotte et al., 2020). If all Fe$^{3+}$ in modern aeolian dust represents iron that would be labile under reducing conditions, this would increase the fraction of dissolvable iron to ~60%, which is the upper limit that we adopt for the Archean and Proterozoic. Note that part of that labile iron could have been leached out and contributed to riverine export rather than continental export (Section 3.2). We assume that the dissolvable fraction was higher in the Archean and Proterozoic than modern day values, corresponding to a lower limit of >10%. The chemical composition of the upper continental crust presumably had a minor role on iron bioavailability as reconstructions of its composition suggest that the Fe concentration of the upper continental crust only changed from ~5 wt% in the Archean to ~3.5 wt% at present due to changes in the nature and the proportions of different igneous rocks (Ptáček et al., 2020).

    Aeolian sediments are found through much of Earth's geological record (Rodríguez-López et al., 2014), with the oldest aeolian quartz sandstones found in the ~3.2 Ga lower Moodies Group of the Swaziland Supergroup in South Africa (Simpson et al., 2012; Eriksson et al., 1998). This shows that some form of wind dust transport was active since at least ~3.2 Ga. In the present contribution, we assume that the flux of aeolian dust from the continents to the oceans simply scales with the areal extent $A(t)$ of emerged continents $\phi_{dust}(t) = \phi_{dust}(t_p) A(t)/A(t_p)$ with $t_p$ denoting present time. The flux of dissolvable iron at any time is therefore,

$$\phi_{Fe,aeo}(t) = \phi_{dust}(t_p) \frac{A(t)}{A(t_p)} [Fe]_{EC}(t)\, F(t),$$

with $[Fe]_{EC}$ the Fe concentration of emerged continents and $F$ the dissolvable fraction of total iron delivered to the oceans. We have $\phi_{dust}(t_p) = 440 \pm 240$ Tg yr$^{-1}$, $A(t)/A(t_p) = 1$ in the Phanerozoic, 1 in the Proterozoic (~1.3 Ga, stage #3), and 0.5 in the Archean (~3 Ga; stage #1) (Korenaga et al., 2017), $[Fe]_{EC}(t) = 3.5$ wt.% in the Phanerozoic, 3.8 wt.% in the Proterozoic, and 5 wt.% in the Archean (Ptáček et al., 2020), and $F(t) = 0.03$ (0.01-0.1) in the Phanerozoic, and 0.1 to 0.6 in the Archean and Proterozoic. The most uncertain term in this assessment is the flux of dust export from the continents to the oceans, as Archean paleogeography and climate are largely unconstrained. With this caveat in mind, we estimate $\phi_{Fe,aeo}$ =**460 [70-2400], 5850 [760-15500], and 3850 [500-10200] Gg yr$^{-1}$ at present, and during stages #3 and #1, respectively** (the numbers in brackets are plausible ranges; the modern value is uncertain because the dissolvable fraction is highly uncertain).

3.2. Riverine transport

Modern estimates of the delivery of dissolved iron by rivers to the oceans suffer from the same difficulties that plague estimates of dissolved iron in the oceans, notably concerning the operational cut-off in grain size (*i.e.*, filter size) used to define dissolved iron (Sections 1 and 2). An additional complication with estimating iron delivery to the oceans is the role played by estuaries and deltas, where mixing between fresh and saline waters induces flocculation and sedimentation of iron-bearing particles that would otherwise have a long residence time in the oceans. "Dissolved" iron in rivers comprises a significant fraction of <0.45 µm negatively charged colloidal particles that get neutralized and flocculate upon mixing with seawater (Boyle et al., 1977). Only a small fraction (~10%) of filterable iron delivered by rivers makes it past estuaries and deltas into the oceans (Boyle et al., 1977). The flux of dissolved (filterable) iron in rivers is ~2470 Gg yr$^{-1}$ (Gaillardet et al., 2003). Assuming 90% removal, this gives a riverine flux past estuaries and deltas of ~247 Gg yr$^{-1}$. This is comparable to the estimate of ~145 Gg yr$^{-1}$ given by DeBaar and De Jong (2001). Tagliabue *et al*. (2016) compared riverine transport of iron in 13 global ocean iron biogeochemistry models. In most of those models, the riverine flux is set to 0. The other 4 models have dissolved iron inputs from rivers of ~4 to 140 Gg yr$^{-1}$. Under the anoxic Archean atmosphere, iron leached from rocks on the continents would have avoided oxidation and would have been a lot more mobile than at present, leading to a higher riverine flux of Fe. To estimate that flux, we use manganese as a proxy (Fig. 8).

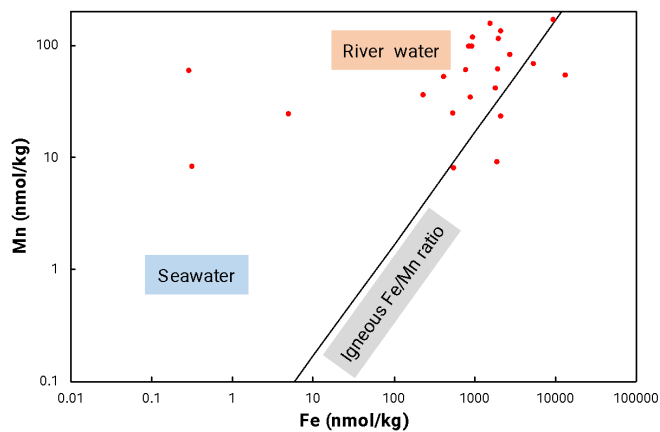

**Fig. 8.** Dissolved iron and manganese concentrations in modern seawater (GEOTRACES, 2021) and major rivers (Gaillardet et al., 2003). The igneous Fe/Mn ratio of ~60 g/g (McDonough and Sun, 1995) is given for comparison. The low Fe/Mn ratio of seawater relative to river waters and igneous rocks is because Fe is more readily oxidized than Mn, limiting its mobility in fluids. To estimate the riverine Fe flux under anoxic conditions, we take the modern riverine Mn flux and multiply it by the igneous Fe/Mn ratio as Fe$^{2+}$ and Mn$^{2+}$ have very similar chemical behaviors.

In both high- and low-temperature systems, Fe$^{2+}$ and Mn$^{2+}$ have very similar behavior. On the present oxygenated Earth, Mn$^{2+}$ leached from igneous rocks can be oxidized into Mn$^{3+}$ and Mn$^{4+}$ that are much less mobile. Compared to Fe$^{2+}$, Mn$^{2+}$ is however less easily oxidized and when it is, it is more readily reduced back to soluble Mn$^{2+}$. This stems from the higher oxidation potential of Mn$^{2+}$/Mn$^{4+}$ than Fe$^{2+}$/Fe$^{3+}$. The higher mobility of Mn relative to Fe on modern Earth is reflected in the dissolved Fe/Mn ratio in rivers (~2 g/g; Gaillardet et al., 2003; ~0.2 g/g after accounting for 90% Fe sequestration in estuaries and deltas; Boyle et al., 1977) compared to igneous rocks (~60 g/g; McDonough and Sun, 1995). Manganese can also be lost by flocculation in estuaries, but it is recycled to the water column by reduction during diagenesis, such that estuarine sediments could even be a net source rather than sink of Mn to the oceans (Sundby et al., 1981). To estimate riverine export of Fe under a globally anoxic atmosphere, we assume that the flux of Mn would remain unchanged, but that the Fe/Mn ratio would be that of igneous rocks, as there would be little oxidative decoupling between $Fe^{2+}$ and $Mn^{2+}$. The modern riverine flux of Mn is 1270 Gg yr$^{-1}$ (Gaillardet et al., 2003; estuaries and deltas are assumed to have a negligible effect on Mn export), which corresponds to an equivalent anoxic Fe export of $1270 \times 60 = 76200$ Gg.yr$^{-1}$.

Other factors could have modulated this flux. Unlike aeolian export, which is associated with physical weathering, riverine transport of iron would have been tied to chemical weathering of the continents, which could have covered a smaller fraction of Earth's surface in the Archean, either because they were covered by seas or had not reached their modern extents (Flament et al. 2008; Bindeman et al. 2018; Albarede et al. 2020). Assuming modern-like $CO_2$-climate feedbacks, at steady-state the rate of silicate weathering would have been tied to the rate of $CO_2$ degassing through the Urey reaction $CaSiO_3(s) + CO_2(g) \rightarrow CaCO_3(s) + SiO_2(s)$, which would have operated even with a smaller fraction of exposed lands, by modulating both precipitation and temperature (Abbot et al., 2012; Graham and Pierrehumbert, 2020; Krissansen-Totton et al., 2018) as well as enhancing submarine weathering of the oceanic crust. On modern Earth, continental weathering represents the principal mechanism for removing $CO_2$, but seafloor weathering could have played a more significant role as a $CO_2$ sink in the past under higher $pCO_2$ atmosphere. Krissansen-Totton et al. (2018) modeled plausible $CO_2$-weathering climate feedback scenarios for Earth's history and concluded that in the early Archean, seafloor weathering may have been a comparable carbon sink to continental weathering (Fig. 9). Taking the median model output, the ratio $r$ of continental/seafloor weathering at 3.1, 1.7, and 0 Ga (present) would have been 4, 13, and 17.

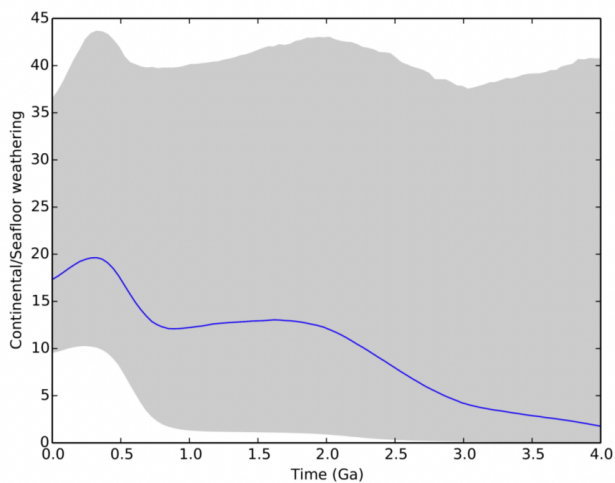

**Fig. 9.** Ratio, $r$, of continental to seafloor weathering through time in the model of Krissansen-Totton et al. (2018).

Chemical weathering of silicates and leaching of Ca on the continents would have been accompanied by Fe mobilization. At steady state, the $CO_2$ degassing flux must be balanced by the weathering flux, much of which happens on the continents. We therefore assume that the riverine export flux of iron simply scales with (*i*) the rate of $CO_2$ degassing through time, which we assume proportional to the rate of heat loss from the Earth's mantle, $\phi_{heat}(t)$ (Thibon et al., 2019), and (*ii*) the fraction $r/(1+r) \simeq 1-1/r$ of $CO_2$ drawdown associated with continental weathering (Krissansen-Totton et al. 2018). Iron mobility will depend on $pO_2$, but parameterizing this influence is difficult, as their relationship is unlikely to be linear. The presence of red beds and iron-rich paleosols starting with the beginning of the GOE suggests that iron mobilization from the continents was already limited since that time and we take the modern as representative of iron riverine export at all times after the GOE (Beukes et al., 2002; Gay and Grandstaff, 1980; Holland, 2006; Rye and Holland, 1998; Yang and Holland, 2003). We thus have for the riverine flux of iron at different times in Earth's history,

$$\phi_{Fe,riv}(t) = \frac{1-1/r(t)}{1-1/r(t_p)} \frac{\phi_{heat}(t)}{\phi_{heat}(t_p)} \phi_{Fe,riv}(t_p), \text{ (\#5 Present, \#3 Proterozoic)}$$

$$\phi_{Fe,riv}(t) = \frac{1-1/r(t)}{1-1/r(t_p)} \frac{\phi_{heat}(t)}{\phi_{heat}(t_p)} \left(\frac{Fe}{Mn}\right)_{igneous} \phi_{Mn,riv}(t_p). \text{ (\#1 Archean)}$$

We have $\phi_{Fe,riv}(t_p) = 247$ Gg yr$^{-1}$; $\left(\frac{Fe}{Mn}\right)_{igneous} \phi_{Mn,riv}(t_p) = 76200$ Gg yr$^{-1}$; $1 - 1/r$=0.75, 0.92, and 0.94 at 3.1, 1.7, and 0 Ga ($t_p$=0 Ga). Establishing how heat loss from the mantle, $\phi_{heat}(t)$, changed through Earth's history is difficult because the relationship that links mantle temperature with convective heat transport is uncertain. A central issue for constraining this relationship is that at present, the convective Urey ratio of the mantle is only ~0.1 to 0.5 (Abe et al., 2022; Jaupart et al., 2007; Korenaga, 2008), meaning that there is a large imbalance between the rate of radiogenic heat production and the rate of heat loss. Models that consider a power-law dependence of heat loss on temperature (higher temperature is associated with lower viscosity and faster convection) yield backward-calculated mantle potential temperatures that are unrealistically high in the Archean, a problem known as the *thermal catastrophe* (Christensen, 1985; Korenaga, 2008). Several solutions have been proposed to solve this conundrum (Butler and Peltier, 2002; Honda, 1995; McKenzie and Richter, 1981; Spohn and Schubert, 1982; Conrad and Hager, 1999; Korenaga, 2006, 2008; Sleep, 2000; Patočka et al., 2020; Stéphane, 2016; Grigné and Labrosse, 2001; Lenardic et al., 2011). We are interested in the mantle heat fluxes at ~3.1 Ga (stage #1), ~1.3 Ga (stage #3), and present (stage #5). Korenaga (2008) argued that the mantle heat flux would not have changed much since 4 Ga. Patočka *et al*. (2020) calculated heat fluxes at 3.1, 1.4, and 0 Ga of ~2.3, 1.5, and 1 times the present-day heat flux, respectively. We take the constant heat flux of Korenaga (2008) and the enhanced heat flux model of Patočka *et al*. (2020) as lower and upper limits on mantle-heat fluxes $\phi_{heat}(t_p)$ as they encompass the values calculated by several other models that can reproduce the convective Urey ratio, plate velocities, and mantle cooling since the Archean. Using the constraints outlined above, we have $\phi_{Fe,riv} = \mathbf{145\ [4-250], 180\ [4\text{-}370],}$ **and $\mathbf{125,000}$ [61,000-182,000] Gg.yr$^{-1}$ at present (stage #5), during stages #3 (Proterozoic) and #1 (Archean), respectively**. While aeolian transport dominates export of dissolved iron from the continents to the oceans at present, our assessment suggests that riverine transport would have dominated continental iron export in the Archean. The calculation of the iron riverine export is surprisingly high, but the approach to derive this value is sound, unless model estimates of the contributions of continental and seafloor weathering under a $CO_2$-rich atmosphere (Krissansen-Totton et al., 2018) are incorrect and seafloor weathering outstripped continental weathering in the Archean eon.

## 4. Hydrothermal sources

## 4.1. Modern hydrothermal sources

Seafloor hydrothermal activity at mid-ocean ridges (MOR) and ridge-flanks is one of the fundamental processes controlling the exchange of heat and chemical species between seawater and oceanic crust (Fig. 10; Edmond, 1979; Stein, 1994; Elderfield, 1993;Wheat, 2000). The interactions between hydrothermal plumes and surrounding seawater might also significantly impact whole-ocean biogeochemical budgets (Kadko, 1993; Elderfield and Schultz, 1996; German and Von Damm, 2003; German and Seyfried, 2014). A major challenge limiting current models of both heat and mass flux through the seafloor is estimating the distribution of the various forms of venting, including focused *vs*. diffuse and ridge axis *vs*. ridge flank. One approach to estimate global hydrothermal fluxes of various chemical elements and species is to combine the seawater flux derived from geophysical constraints with the composition of hydrothermal fluids at various temperatures. Conversion of the hydrothermal power outputs through thermal calculation to hydrothermal water and chemical fluxes requires assumptions regarding heat loss from ridge axis (crustal ages of 0 - 0.1 Ma), near-axial settings (with crustal ages of 0.1 - 1 Ma), and ridge flanks (> 1 Ma) (Mottl and Wheat, 1994).

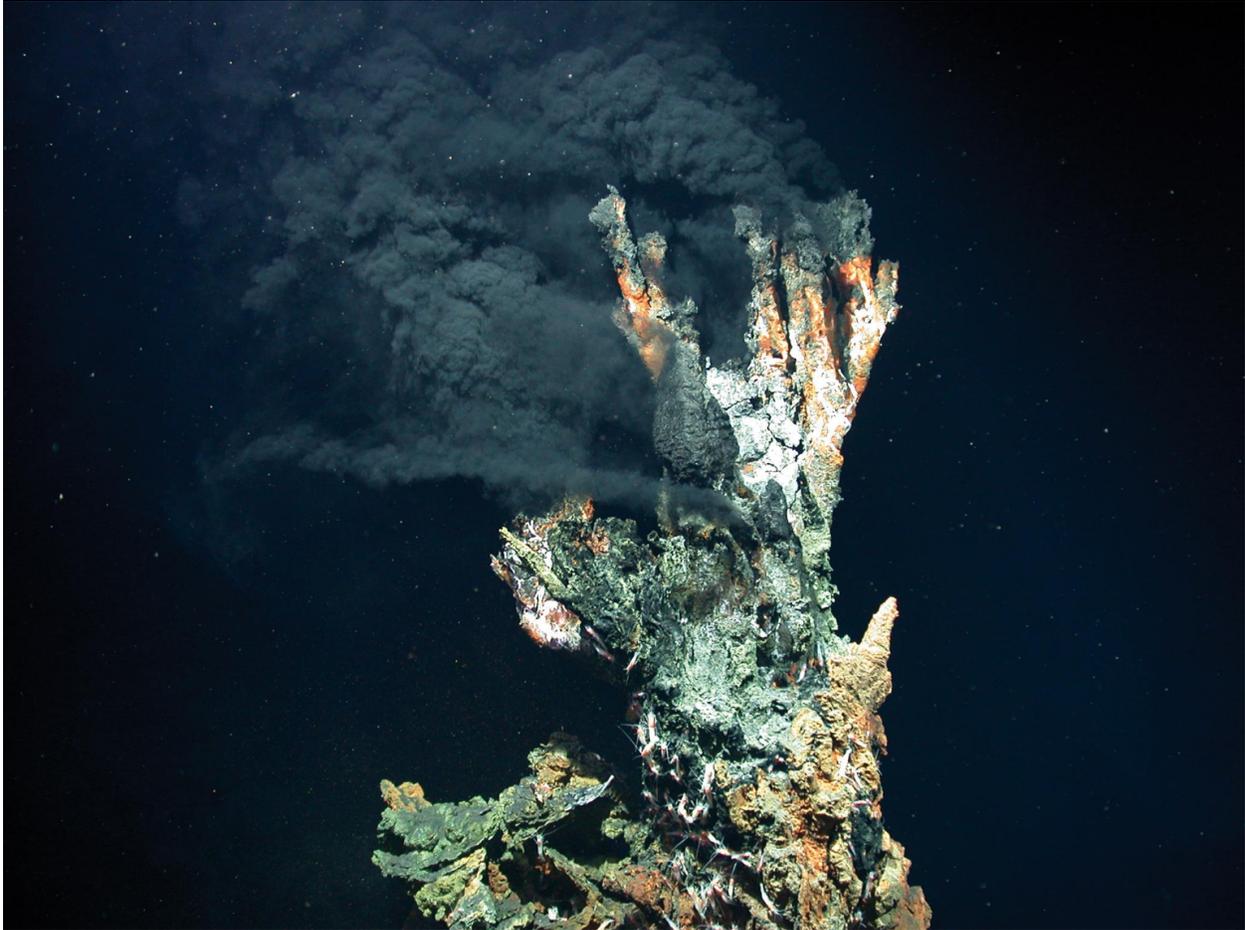

**Fig. 10.** Iron-rich fluids from a "black smoker" chinney at a depth of 3,300 m in the Logatchev hydrothermal field on the Mid-Atlantic Ridge (Credit: MARUM – Center for Marine Environmental Sciences, University of Bremen, CC-BY 4.0).

*4.1.1. Axial and near-axial settings*

The best estimates of the global power output at mid ocean ridges (*i.e.*, crust 0–1 Ma in age) range from $1.50 \pm 0.18$ TW to $3.3 \pm 0.6$ TW (Mottl, 2003). The difference is mainly related to spreading rates, with lower limits for fast-spreading ridges. Using this approach, the likely power output for crust 0–1 Ma in age is $2.8 \pm 0.4$ TW (Mottl, 2003). The best estimate for the global hydrothermal power output at the axis (0 - 0.1 Ma in age) is $1.8 \pm 0.3$ TW. The hydrothermal power output in the near-axial region, through crust between 0.1 and 1 Ma in age, can thus be obtained by subtracting the output at the axis from the total output; $2.8 \pm 0.4 - 1.8 \pm 0.4 = 1.0 \pm 0.5$ TW. How the axial and near-axial power output values translate into fluxes of heated seawater and dissolved chemicals depends mainly on the temperature at which this heat is removed. For an exit temperature of $350 \pm 30°C$ and the associated enthalpy of $1540 \pm 200$ J/g at 300–600 bars, the minimum seawater flux required to remove the minimum power output of $1.80 \pm 0.3$ TW is $3.70 \pm 0.51 \times 10^{16}$ g/yr, a rate at which the entire mass of the oceans ($1.41 \pm 10^{24}$ g) would cycle through the crust in 38 million years (Mottl, 2003). For a total power output of $2.8 \pm 0.3$ TW, including near axial settings, the corresponding seawater flux would be $5.9 \pm 0.8 \times 10^{16}$ g/yr.

The determination of high temperature hydrothermal Fe fluxes also requires the assessment of the average concentrations of Fe in hydrothermal fluid endmembers, which vary depending on venting temperature, salinity, and geological settings. Here, we determine the global average using a recent compilation of > 1000 discrete vent fluid datapoints (Diehl and Bach, 2020) from worldwide vent sites. As a first-order approach, we calculated the average Fe concentration using the entire dataset, which yielded large uncertainties. The data were therefore corrected for seawater mixing (end-member values at zero Mg concentrations; Edmond et al., 1979a), and the resulting database was then filtered for temperature > 340°C to avoid the effect of subsurface Fe precipitation upon cooling of the hydrothermal fluids and phase separation. Although the chlorinity of hot springs varies widely, nearly all the reacted fluid, whether vapor or brine, must eventually exit the crust within the axial region. Steady state mass balance considerations on seawater salinity dictates that the integrated hotspring flux must therefore have a chlorinity like that of seawater. Using this approach, we determined an Fe concentration of $5.8 \pm 3.4$ mmol/kg, which is the average concentration in hydrothermal vent fluids that have an exit temperature > 340°C and seawater-like salinity. The significant range of Fe concentrations in high-temperature hydrothermal fluids likely reflect the range of geological settings (fast- vs. slow-spreading ridges) and host-rock composition (ultramafic, basaltic or felsic). In particular, ultramafic-hosted vents show the highest Fe concentrations at > 10 mM, but their overall significance remains poorly constrained.

Using the high temperature hydrothermal water flux of $5.9 \pm 0.8 \times 10^{16}$ g/yr estimated above, and our new estimate of average Fe concentration, we calculated a high-temperature hydrothermal Fe flux of **$0.34 \pm 0.22 \times 10^{12}$ mol/yr ($19000 \pm 12000$ Gg.yr$^{-1}$)**. This is within the high end of previous estimates by Elderfield and Schultz (1996) and similar to more recent estimates (Poulton and Raiswell, 2002; Thompson et al., 2019).

High-temperature hydrothermal fluids may, however, not be entirely responsible for the transport of all the axial hydrothermal heat flux. Elderfield and Schultz (1996) considered a uniform distribution, on the global scale, in which only 10% of the total axial hydrothermal flux corresponded to 'focused' flow (heat flux 0.2–0.4 TW; water flux 0.3–0.6 $10^{16}$g/yr). A more recent calculation of the high-temperature water flux based on the mass balance of Tl in the oceans yields a value of 0.17–2.93 $10^{16}$g/yr (Nielsen et al., 2006). This suggests that only 10 to 20% of the thermal energy is released at mid-ocean ridge axes (Elderfield and Schultz, 1996; Nielsen et al., 2006) with the remainder of the axial heat flux likely transported by a much larger volume flux of lower temperature fluid. Because Fe concentrations in diffuse hydrothermal fluids are affected by subsurface Fe precipitation during cooling of the hydrothermal fluid (through Fe sulfide precipitation), the global hydrothermal Fe flux is probably affected by the nature (focused *vs*. diffuse) of the axial fluid flow. The estimated axial hydrothermal Fe flux of $0.34 \times 10^{12}$ mol/yr should therefore be considered as a maximum value.

*4.1.2. Ridge flank settings*

Newly formed oceanic crust undergoes extensive chemical and physical alteration driven by circulation of seawater for at least 65 Myr, and this interaction exerts a major control on the chemistry of many elements in seawater. For the transport of heat and many components such as the alkali elements, Ca, U, Mg, and C, seawater circulation in ridge flanks plays a role rivaling that of axial hydrothermal systems (Mottl and Wheat, 1994; Wheat and Mottl, 2000). The total hydrothermal power output through ridge flanks, from 1 Ma to a sealing age of 65 Ma, has been estimated at $7.1 \pm 2$ TW. The major difficulty with estimating ridge flank geochemical fluxes is that the amount of heat loss at a particular temperature and, thus, the amount of the seawater flux at that temperature are unknown. This difficulty is compounded by the scarcity of fluid data and non-conservative behavior of Fe with respect to secondary mineral precipitation, either as $Fe^{3+}$ or $Fe^{2+}$ species.

Given that most of the ridge-flank hydrothermal power output should occur at cool sites (< 20°C), the flux of slightly altered seawater could range from 0.2 to $2 \times 10^{19}$ g/yr, which is within the same order of magnitude as the global flux of river water to the ocean of $3.8 \times 10^{19}$ g/yr (Mottl, 2003). Whether ridge flanks are important for global Fe flux depends on how much the seawater changes in composition as it circulates through the crust as only small changes are needed in the composition of such a large water volume for significant Fe fluxes to potentially occur. .

Wheat *et al*. (2002) analyzed hydrothermal fluids derived from 3.5 Ma basaltic crust on the eastern flank of the Juan de Fuca Ridge and found Fe concentrations below the detection limit of 0.05 umol/kg, suggesting that Fe is efficiently precipitated during basement alteration at ridge flanks, presumably as $Fe^{3+}$ oxyhydroxides and/or Fe-sulfides. This is consistent with the exhaustive compilation of $Fe_2O_3$, FeO, and S concentrations from DSDP/ODP drill core samples representing upper basaltic ocean crust. The latter data were used to infer that $Fe^{3+}$/total Fe increased from $0.15 \pm 0.05$ to $0.45 \pm 0.15$ within the first 10–20 Myr of crustal evolution (Bach and Edwards, 2003). Assuming a seafloor production rate of $3.0 \pm 0.5$ km$^2$/yr, a depth extent of oxidation of $500 \pm 200$ m, an upper crustal porosity of $0.10 \pm 0.05$, and a basalt density of $2950 \pm 50$ kg/m$^3$, an upper crustal production rate of $4.0 \pm 1.8 \times 10^{12}$ kg/yr can be calculated. Using the Fe oxidation budget estimated above, Bach and Edwards (2003) calculated a global $Fe^{2+}$ oxidation rate of **$1.7 \pm 1.2 \times 10^{12}$ mol Fe/yr ($95000 \pm 67000$ Gg.yr$^{-1}$)**. This flux is by itself larger than the high temperature hydrothermal flux, but is unlikely to be contributing significantly to the oceanic budget considering the low solubility of Fe under oxygenated seawater conditions. This situation may be however different under anoxic conditions relevant to Precambrian oceans, as this Fe pool could be mobilized in low-temperature fluids and contribute to oceanic budget as seawater circulates through ridge flanks and Fe is transported away from the deep sea.

We estimated the potential of ridge-flank fluid flow to contribute to oceanic Fe using a previously determined Si flux, and assuming a constant Fe/Si ratio in warm ridge flank fluids. Because Si in hydrothermal fluids behaves mostly as a conservative element during mixing with oxygenated seawater, this approach should provide a means of estimating the maximum Fe flux mobilized under modern conditions. This is an upper limit as some Fe released in the deep sea with Si could be locally oxidized and would therefore not participate in the global biogeochemical cycle of iron. Silicon fluxes related to seawater-crust exchange at ridge flanks has been previously determined through direct monitoring of fluids produced in low temperature hydrothermal circulation (Wheat and Mottl, 2000). Wheat and McManus (2005) determined a global Si flux of $0.011 \times 10^{12}$ mol/yr for warm ridge flank fluids. This estimate is based on the measured Si anomaly associated with a warm spring (0.17 mmol/kg) and a ridge flank fluid flux (equivalent to 0.5 to 1 $\times 10^{18}$ g/yr) determined using oceanic Mg mass balance, assuming that the ocean is at steady state with respect to Mg. A more recent study, based on analysis of fluid compositions in cold and oxygenated ridge flank settings (*e.g.*, North Pond, Mid-Atlantic Ridge) confirms that incipient alteration of volcanic rocks may result in significant release of silica (0.07 mmol/kg water) to circulating seawater (Meyer et al., 2016). As a first approximation, we used an average Si/Fe ratio of 5.2 ± 0.5 mol/mol determined from the extensive dataset of warm hydrothermal fluids from Kama'ehuakanaloa (formerly Lō'ihi) Seamount (Glazer and Rouxel, 2009). This value compares well with the average Si/Fe ratio in high temperature vent fluids (> 340°C) determined at 7.1 mol/mol (and ranging from <0.1 to 20 mol/mol). Using this approach, we determined a low temperature off-axis Fe flux of **$2.1 \times 10^9$ mol/yr (120 Gg.yr$^{-1}$)**. This estimate is several orders of magnitude lower than the rate of oceanic crustal Fe oxidation of **$1.7 \pm 1.2 \times 10^{12}$ mol Fe/yr (95000 ± 67000 Gg.yr$^{-1}$)**, calculated above, suggesting that much of the iron released in the fluid in ridge-flanks is immobilized by oxidation and precipitation of secondary minerals. It is also insignificant compared to high-temperature fluxes.

*4.1.3. Comparison with Thompson et al. (2019)*
Thompson *et al*. (2019) recently re-estimated the modern hydrothermal Fe flux using the study of Halevy and Bachan (2017). Thompson *et al*. (2019) determined a modern on-axis hydrothermal Fe$^{2+}$ flux of $0.30 \times 10^{12}$ mol/yr (17000 Gg.yr$^{-1}$) identical to our inferred axial hydrothermal Fe$^{2+}$ flux of $0.34 \pm 0.22 \times 10^{12}$ mol/yr (19000±12000 Gg.yr$^{-1}$). In marked contrast, however, Thompson *et al*. (2019) determined a modern off-axis hydrothermal Fe$^{2+}$ flux of **$10.5 \times 10^{12}$ mol/yr (590000 Gg.yr$^{-1}$)**, 3 orders of magnitude higher than our estimate at **$2.1 \times 10^9$ mol/yr (120 Gg.yr$^{-1}$)**. This large discrepancy arises from the use of an overestimated Fe$^{2+}$ concentration of 0.75 mmol/kg for off-axis fluids. Thompson *et al*. (2019) cite the study of Poulton and Raiswell (2002) as justification for adopting this value. However, a concentration of 0.75 mmol/kg represents the lower end for high temperature, on-axis hydrothermal fluids used by Elderfield and Schultz (1996) rather than off-axis hydrothermal fluid. Poulton and Raiswell (2002) cited the study of Mottl et al. (1998) who found 0.34 ppm Fe (6 umol/kg) in the warm springs emitted from basaltic outcrops on the flank of the Juan de Fuca Ridge. These authors revised this concentration anomaly in warm fluids as being insignificant relative to measured background seawater (Wheat and Mottl, 2000) and assumed a net ridge-flank Fe$^{2+}$ flux of close to zero. This assumption is further confirmed by a more recent study of colder (< 25°C) ridge flank settings from slow-spreading ridges, which yield even lower Fe$^{2+}$ than warm ridge flanks, due to the oxic nature of circulating fluids (Orcutt et al., 2013). We therefore stand with a net ridge-flank Fe input being at a maximum of **$2.1 \times 10^9$ mol/yr (120 Gg.yr$^{-1}$).**

**4.2. Importance of hydrothermal plume processes**

When metal and sulfide-rich hydrothermal fluids emerge from seafloor vents at temperatures up to 350°C, the first mineral phases that rapidly begin to precipitate within a few cm and less than one second (Rudnicki and Elderfield, 1993) are fine-grained Fe-sulfide minerals, dominantly pyrrhotite (Mottl and McConachy, 1990), which create characteristic 'black smoker' plumes (Von Damm, 1990). On average, approximately 50 % of the initial dissolved $Fe^{2+}$ is removed early by local precipitation of Fe sulfides, while most of the remaining $Fe^{2+}$ in the buoyant plume undergoes oxidation and precipitation as particulate $Fe^{3+}$ oxyhydroxides (Baker and Massoth, 1987; Field and Sherrell, 2000; Mottl and McConachy, 1990; Rudnicki and Elderfield, 1993; Statham et al., 2005; Trocine and Trefry, 1988). Rates of $Fe^{2+}$ oxidation in the buoyant plume that drive the precipitation of $Fe^{3+}$ oxyhydroxide particles are highly variable, with half-lives ranging from 17 min at the Mid-Atlantic Ridge above the Transatlantic Geotraverse (TAG) site, to 6 hr at the Juan de Fuca Ridge in the Northeast Pacific (Baker and Massoth, 1987; Field and Sherrell, 2000; Mottl and McConachy, 1990; Rudnicki and Elderfield, 1993; Statham et al., 2005; Trocine and Trefry, 1988). Rates of V oxidation in the buoyant plume appear to be dominantly controlled by the geochemistry of local seawater, particularly pH and dissolved $O_2$, rather than aspects of vent fluid chemistry such as the $Fe^{2+}$ content or Fe/S ratio (Field and Sherrell, 2000), reflecting the major impacts of $OH^-$ and $O_2$ on $Fe^{2+}$ oxidation kinetics (Millero et al., 1987; Sung and Morgan, 1980). This is clearly demonstrated by increasing $Fe^{2+}$ oxidation half-lives around the ocean conveyor belt, with the slowest oxidation occurring in the Pacific Ocean where pH and $O_2$ are lower than in the Atlantic Ocean, with the intermediate composition of the Indian Ocean driving intermediate $Fe^{2+}$ oxidation rates (Statham et al., 2005).

Because it was commonly considered that hydrothermally-released metals were largely removed from seawater by precipitation of iron-bearing minerals within plumes and then deposited at the seafloor close to vent sites, the contribution of hydrothermal fluxes to the ocean budget was long considered small to negligible. Subsequent work, however, questioned the validity of that assumption in what is known as the "leaky vent" hypothesis (Toner et al., 2012). In particular, research in near-vent settings have revealed several potential chemical forms for the delivery of hydrothermally-sourced metals to open oceans: (*i*) complexes with dissolved and particulate organic carbon (DOC and POC) (Bennett et al., 2008; Toner et al., 2009; Sander and Koschinsky, 2011; Hawkes et al., 2013); (*ii*) nanoparticulate minerals, in particular pyrite (Yucel et al., 2011; Gartman et al., 2014 Findlay et al., 2019) and colloidal $Fe^{3+}$ oxyhydroxides (Tagliabue and Resing, 2016); (*iii*) uptake by water column microorganisms (Li et al., 2014). Previous works also raised fundamental questions regarding the residence time of hydrothermal particles in seawater and their dissolution kinetics (Feely et al., 1987; Lilley et al., 1995; Adams et al., 2011; Sylvan et al., 2012; Carazzo et al., 2013). Stabilized hydrothermal Fe from 'leaky vents' to the deep ocean Fe pool could be a globally significant source (Toner et al., 2012), with Bennett et al. (2008) estimating that the ~4 % of hydrothermal Fe stabilized by the available ligands cycling through hydrothermal vents could supply 11-22 % of the global deep ocean dissolved Fe pool. In our treatment of the high temperature axial flux, extrapolation of the 4 % stable export of Fe to the oceans results in a global flux of **760±480 Gg/yr (0.04 × 19000±12000 Gg/yr)**. Research in far-field vent settings have highlighted clear hydrothermal contributions for Fe, Mn, and probably other metals. Long-range (thousands of km) transport of hydrothermal Fe has been revealed by mid-water column Fe "anomalies" in (*i*) the central Pacific ocean (Boyle et al., 2005; Jenkins et al., 2020; Wu et al., 2011); (*ii*) the Southern and Arctic oceans (Fitzsimmons et al., 2017, 2014b; Klunder et al., 2012; Resing et al., 2015); (*iii*) the equatorial Atlantic ocean (Saito et al., 2013); and (*iv*) the Indian ocean (Nishioka et al., 2013). As a consequence, hydrothermal plume processes should be considered in order to determine the global contribution of hydrothermal Fe to the open ocean (Tagliabue et al., 2010).

## 4.3. Precambrian hydrothermal flux

Release of hydrothermal fluids is likely to be related to the rate of heat loss from oceanic crust (Lowell and Keller, 2023; Thompson et al., 2019; Thibon et al., 2019). Lowell and Keller (2023) made the case that hydrothermal heat loss should scale as the square of global lithospheric heat loss. While heat loss from the oceanic crust is relatively well constrained on modern Earth, estimating oceanic heat flux in the Precambrian is not straightforward because simple scaling of viscosity-limited mantle heat loss with temperature yields temperatures in the Archean that are unrealistically high (Christensen, 1985; Korenaga, 2008). Solutions to this problem are discussed in Sect. 3.2 and give heat fluxes that are nearly constant or decrease through time to the present-day value. We take $\frac{\phi_{heat}(t)}{\phi_{heat}(t_p)}$ =1 at present (stage #5), 1 to 1.5 between the GOE and NOE (stage #3), and 1 to 2.3 before the GOE (stage #1). The hydrothermal heat (and Fe flux) could therefore have been higher by factor of 1 to 2 ($1^2$ to $1.5^2$) in stage #3 and 1 to 5 in stage #1 ($1^2$ to $2.3^2$), relative to the present. Other parameters would have likely influenced the hydrothermal flux of iron from altered oceanic crust, most notably the compositions of the rocks (higher degree partial melting during the Archean would have produced more Mg-rich lavas) and fluids (Archean seawater would have contained little sulfate and a significant amount of $CO_2$). Additionally, the anoxic, sulfate-poor conditions of Precambrian seawater (Kump and Seyfried, 2005) through much of early Earth history should have dramatically impacted the efficiency with which hydrothermal fluid Fe was dispersed throughout the global ocean. Rather than rapidly, near-quantitatively precipitating from plumes close to vents as voluminous oxide and sulfide sediments, a lack of bottom water $O_2$ or large seawater sulfate pool from which to generate abundant $H_2S$ may have allowed long-distance transport of a large fraction of hydrothermal $Fe^{2+}$.

The luminosity of the Sun increased through Earth history, which assuming a $CO_2$-dominated climate feedback requires that the $pCO_2$ decreased through time. It is estimated that $pCO_2$ was $3\times10^{-4}$ bar in pre-industrial times (stage #5), $3\times10^{-3}$ bar [$0.6\times10^{-3}$ to $30\times10^{-3}$] at 1.3 Ga (stage #3), and $4\times10^{-2}$ bar [$0.5\times10^{-2}$ to $20\times10^{-2}$] at ~3.1 Ga (stage #1) (Catling and Zahnle, 2020). These high partial pressures would mean that significant dissolved $CO_2$ would have been present at the ingress of hydrothermally circulating fluids. While Phanerozoic $pCO_2$ reconstructions largely overlap with estimates for the Proterozoic, there is little doubt that $pCO_2$ in the Archean was much higher than what it is today. Experiments show that higher $CO_2$ could have a significant effect on the hydrothermal flux of iron (Shibuya et al., 2010; Shibuya et al., 2013; Ueda et al., 2016; Ueda et al., 2021). Alteration experiments conducted at 250 and 350°C with $CO_2$ on Mg-rich basalt similar in composition to Archean seafloor show that under these conditions, the pH of the fluid is raised from 6.5 before water-rock interaction to 6.6-7.2 after (Shibuya et al., 2013) due to basalt carbonation. For comparison, the pH of high temperature hydrothermal fluids at present is ~5. Under Archean conditions, experiments show that mobilization of iron from basalt could have been significantly reduced (Shibuya et al., 2013). Shibuya *et al*. (2013) reported Fe concentrations of 0.002 to 0.042 mmol/L in Archean-like seawater fluids after interaction with basalt. For comparison, modern on-axis high temperature hydrothermal fluids have Fe concentrations of ~6 mmol/L (Sect. 4.1.2). Iron mobilization in low T hydrothermal settings would likely be even smaller (Ueda et al., 2021). Shibuya *et al*. (2010; 2013) argued that hydrothermal circulation associated with extensional settings could have been a net sink for iron rather than a source in the Archean, and they suggested that oceanic plateau and island arc settings could have been more important contributors to the inventory of iron in the ocean.

Higher mantle potential temperature in the Archean means that ultramafic magmas, akin to komatiites, could have been formed at mid ocean ridges and subjected to hydrothermal circulation. Hydrothermal alteration of such rocks was studied by Shibuya *et al*. (2015) and Ueda *et al*. (2016; 2021). They found that in low-T ultramafic systems, Fe mobilization would be small because Fe would be retained as carbonate, but in high-T systems, some Fe could have been released, with the fluids at 300 °C containing up to 0.6 mmol/L of dissolved Fe.

As outlined above, estimating the hydrothermal flux of Fe in the Archean is fraught with difficulties. Experimental results could be taken as evidence that under a high $CO_2$ atmosphere, Fe was not mobile in hydrothermal fluids, but this contradicts geochemical evidence that rare earth elements and by inference Fe in banded iron formations had a significant mantle hydrothermal contribution (Derry and Jacobsen, 1990; Hu et al., 2020; Jacobsen and Pimentel-Klose, 1988; Wang et al., 2016), with the caveat that some banded iron formations could have also been influenced by a flux derived from the continental crust (Alexander et al., 2008; Alibert and McCulloch, 1993; Li et al., 2015). The issue with interpreting this record is that banded iron formations may represent episodic depositional events associated with the emplacement of large igneous provinces (LIPs; Bekker et al., 2010) and the record that they provide might not be representative of the steady state periods of Earth's history that we are targeting here. Another possibility is that high-T alteration of komatiitic magmas could have been a source of Fe in the Archean (Shibuya et al., 2015; Ueda et al., 2016; Ueda et al., 2021). Finally, the origin of hydrothermal Fe might need to be sought elsewhere in environments that lack analogs in the modern world (Shibuya et al., 2010; Shibuya et al., 2013).

To estimate the maximum possible hydrothermal flux of Fe in the Archean and Proterozoic, we take the total modern Fe hydrothermal flux **19000 ± 12000 Gg yr$^{-1}$** before plume processing and use the multiplication factors of 1.5±0.5 and 3±2 discussed above for the Proterozoic and Archean (Lowell and Keller 2003) . We therefore obtain maximum hydrothermal iron fluxes of **28000 ± 20000 to 57000 ± 50000 Gg yr$^{-1}$**. For comparison, Lowell and Keller (2003) calculated a hydrothermal Fe flux of ~2000 Gg/yr between 2.7 and 1.8 Ga. This is an order of magnitude different from our estimate and the discrepancy seems to stem primarily from the value of the modern high-T hydrothermal Fe flux adopted.

For establishing a lower bound, we use the experimental results of Shibuya *et al*. (2013) for 250-350 °C hydrothermal alteration of basalt under the high $pCO_2$ Archean atmosphere, which give an average Fe concentration in the fluid of ~0.01 mmol/L, compared to ~6 mmol/L in high-T hydrothermal fluids at present. Since we want to set the lower bound for the hydrothermal flux in the Archean, we assume that the mantle heat flux did not change through time (Korenaga, 2008) and scale the modern hydrothermal flux by the Fe concentrations obtained in laboratory experiments (Shibuya et al., 2013), resulting in an Archean minimum Fe flux of only 30 Gg.yr$^{-1}$.

Constraints on the metal export efficiency of Archean hydrothermal systems are generally lacking. However, Jamieson et al. (2013) estimated a similar seafloor metal deposition efficiency within the hydrothermal mound of the ~2.7 Ga Kidd Creek seafloor volcanogenic massive sulfide deposit as in modern systems including the Mid-Atlantic TAG mound, such that differences in plume processes should be the main driver for contrasting Fe export efficiencies in ancient *vs*. modern hydrothermal vents. We can reasonably assume an upper limit efficiency of 100%, due to the lack of bottom water $O_2$ (Stolper and Keller, 2018) and low [$H_2S$] in vent fluids that would inhibit voluminous, exhalative, sulfide precipitation (Kump and Seyfried, 2005). Thus, the Archean hydrothermal Fe flux to the global oceans was likely within the same order of magnitude as the primary fluid flux ranges estimated immediately above.

For the Proterozoic, it is difficult to estimate the minimum Fe flux, because significant portions of the deep ocean may have been euxinic (Lyons et al., 2009), with widely developed ferruginous (Planavsky et al., 2011), suboxic (Slack et al., 2007, 2009), and even possibly oxic (Slotznick et al., 2019) conditions, and $pCO_2$ estimates largely overlap with those for both modern and Archean atmospheres (Catling and Zahnle, 2020). Ancient deep ocean redox conditions must have played a major role in determining the efficiency of hydrothermal Fe export from vents to the global oceans, and these are highly uncertain throughout the Proterozoic (Lyons et al., 2014). Whether Fe from hydrothermal plumes was precipitated as oxides (in oxic oceans), sulfides (in euxinic oceans and by reaction with hydrothermal $H_2S$), or transported over long distances as soluble $Fe^{2+}$ (in ferruginous oceans), and in what proportions, remains an open question. Conservatively, we assume that the Fe hydrothermal flux in the Proterozoic was above the lowest estimate for the Archean (>30 $Gg.yr^{-1}$).

## 5. Sedimentary sinks and sources

Sediments can have an ambivalent role, as they are both sinks and sources. In the modern day, sediments represent a major source of iron to the oceans, as highly-reactive ferric iron-bearing minerals in sediments can be cycled back into porewaters and the water column through dissimilatory iron reduction (DIR), which involves the respiration of organic carbon using ferric iron as an electron acceptor. The role of DIR under an anoxic atmosphere should have depended strongly on the availability of ferric oxide substrates. The nature and fluxes of sedimentary iron sinks could have also changed dramatically in the past, responding to changing seawater compositions. Below, we start by discussing sedimentary iron sinks (oxides and sulfides) and then discuss sedimentary iron sources, as they involve the recycling of sedimentary sinks in the water column.

### 5.1 Iron formations and associated minerals

Iron-bearing minerals such as Fe oxides are commonly found interbedded with silica-rich layers forming extensive deposits throughout the Precambrian (Fig. 11). These chemical sedimentary archives, also known as iron formations (IFs), represent a dominant Fe sink during the Archean and early Proterozoic eons (James, 1954; Cloud, 1973; Garrels et al., 1973). Since the 1950s, a vast body of work has been published focused on the origin of IFs, including laboratory studies aimed at reproducing their precipitation mechanisms and conditions, and analyses of natural samples (Bekker et al., 2010, 2014; James, 1954; Trendall and Blockley, 1970; Cairns-Smith, 1978; Anbar and Holland, 1992; Isley, 1995; Konhauser et al., 2002; Dauphas et al., 2004; Posth et al., 2008; Planavsky et al., 2010a; Köhler et al., 2013; Rasmussen et al., 2014; Tosca et al., 2016; Halevy et al., 2017; Nie et al., 2017; Rouxel et al., 2018; Thompson et al., 2019; Rego et al., 2021; Dodd et al., 2022). Despite these efforts, many aspects of the origin of banded iron formations remain uncertain.

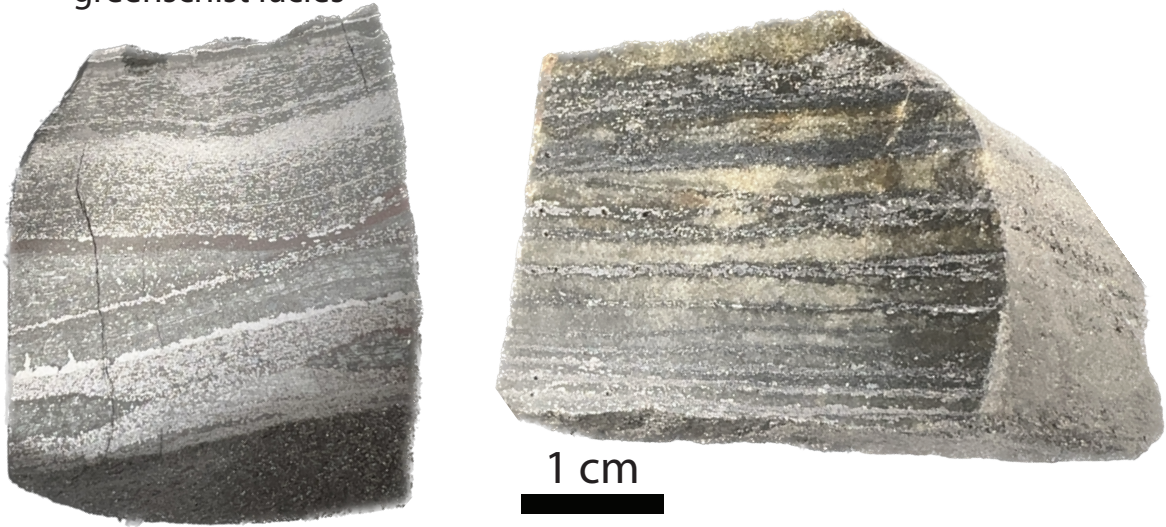

***Fig. 11.*** *Examples of BIFs from the ~2.50-2.45 Ga Dales Gorge formation, Hamersley, Australia (left) and ~3.8-3.7 Isua Supracrustal Belt (right). The Isua sample (IS-04-06) was studied by Dauphas et al. (2007). The layers are composed primarily of magnetite intercalated with quartz.*

In general, there is a common view that ferrous iron ($Fe^{2+}$), which is highly soluble in anoxic conditions, is oxidized to form insoluble ferric iron ($Fe^{3+}$) minerals that are common in IFs. One interpretation is that mobile dissolved $Fe^{2+}$ was abundant in reducing oceans, particularly during the first half of Earth's history. It was transported over great distances until it reached surface marine environments, where $Fe^{2+}$ was fully and/or partially oxidized to form insoluble $Fe^{3+}$ or mixed valence iron (oxyhydrox)oxides. These precipitates would settle to the ocean floor along with amorphous silica gels, forming deposits consisting of alternate bands of Fe- and Si-rich layers. Iron oxidation could have been mediated by oxygen produced by cyanobacteria (Cloud, 1973), anoxygenic photosynthesis whereby carbon reduction is coupled to iron oxidation (Widdel et al., 1993; Kappler et al., 2005; Ozaki et al., 2019; Thompson et al., 2019; Rego et al., 2021), and/ or oxidation via UV radiation (Cairns-Smith, 1978; Braterman et al., 1983; Nie et al., 2017). Alternatively, other studies have suggested that an Fe-silicate mineral, greenalite [$(Fe^{2+}, Fe^{3+})_{2-3}Si_2O_5(OH)_4$], was the initial precursor of IFs (Rasmussen et al., 2021). In this interpretation, ferric minerals are the product of a secondary oxidation process. Regardless of the mineralogical precursor for IFs, there is a consensus that a large amount of Fe was removed from seawater as Fe-mineral precipitates, which led to the formation of world-class iron ore deposits we observe today.

The iron formation Fe sink (*i.e.* the amount of Fe removed from Earth's oceans through time as chemical precipitates) can be estimated by means of the available geological record. A difficulty with this approach is the preservation bias of the rock record. An additional complication is that many iron formations do not have well-constrained geochronological data and detailed paleogeographic reconstructions containing information on stratigraphic thickness and lateral extent. In well-studied locations, such as the Hamersley basin, chemical sediment accumulation rates were calculated to range between 3 to 15 m.Myr$^{-1}$ (Lantink et al., 2022), which are similar to values obtained from older IF deposits on the Amazonian craton (*e.g.* 6 m.Myr$^{-1}$, Carajás basin; Rossignol et al., 2023). Assuming that the sedimentation rate did not change significantly through time, we can calculate the flux of Fe accumulating in iron formation $i$ ($\phi_{Fe-IF,i}$) by using available information on each deposit's areal extent ($S_i$ in km$^2$), Fe concentration ($[Fe]_i$ ~24-39 wt.%), density ($\rho$~3.5 g/cm$^3$; Konhauser et al., 2018), and deposition rate of iron formation ($r_i$ in m/Myr) (Table 1),

$$\phi_{Fe-IF,i} = \frac{\rho S_i h_i [Fe]_i}{h_i/r_i} = \rho S_i r_i [Fe]_i,$$

where $h_i$ is the unit thickness, which does not play a role in the calculation of Fe deposition rate. We use the average Fe$_2$O$_3$ concentration of IF from each locality, which varies between 32.4 to 50 wt.% to calculate Fe content in IF and consider 3 deposition rates of 6, 15, and 50 m/Myr. We compiled the aerial extents of major iron formations, allowing us to calculate an estimate for the Fe deposition flux for each episode of IF deposition, which is plotted in Fig. 12 assuming a deposition rate of 15 m/Myr. We also calculate the time span for each IF deposition event. The red bars in that diagram represent each IF deposition episode, while the black continuous line represents the sum of all contributions when IFs overlap in age. As shown, the rate of IF deposition was low before 3 Ga and it saw its maximum immediately before the GOE, defining an average Fe deposition rate in IFs of **102 $\pm$ 139 Gg.yr$^{-1}$** between ~3.8 and 2.4 Ga. IFs almost disappear from the geological record after this period. The estimates given by enumerating all IF occurrences in the geological record provides a minimum estimate of the Fe deposition rate because many IF occurrences could have been eliminated from the record by subsequent geological processes. Accordingly, we estimate minimum IF deposition rates of **102 $\pm$ 139, 44.73, and 1.28 Gg.yr$^{-1}$ during stages #1, #3, and #5, respectively**.

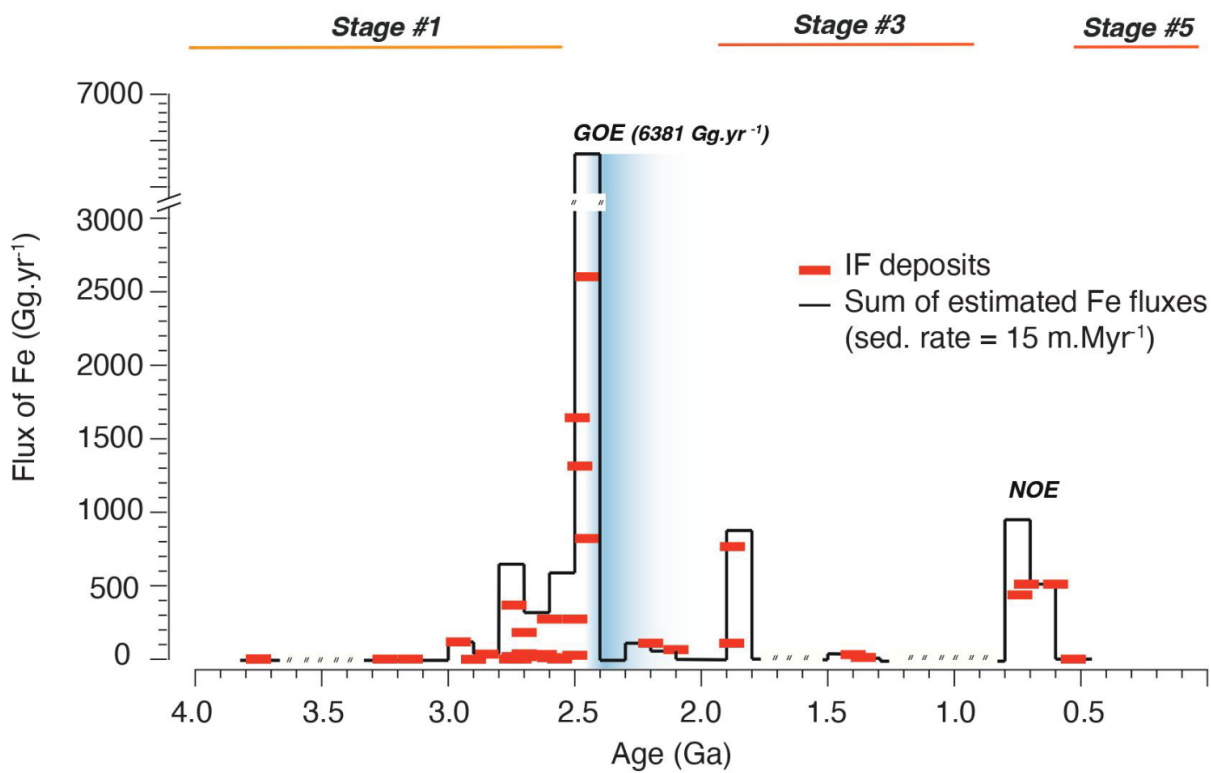

**Fig 12.** Minimum flux of iron into IFs calculated based on the known time span of deposition of IFs, their areal extent, and assuming a deposition rate of 15 m/Myr. Each red segment represents an IF. The black solid line adds up all IF fluxes calculated using a bin of 0.1 Ga, ranging from 4 to 0.5 Ga.

**Table 1**. Calculated minimum, medium and maximum fluxes based on varying depositional rates (6 m.Myr$^{-1}$, 15 m.Myr$^{-1}$, and 50 m.Myr$^{-1}$, respectively) from iron formations deposited within stages 1, 3 and 5.

| | | Mass of Fe accumulated per year | | |
| --- | --- | --- | --- | --- |
| Age | IF Deposits | Min.[1] | Med.[2] | Max.[3] |
| (Ga) | | (Gg.yr$^{-1}$) | (Gg.yr$^{-1}$) | (Gg.yr$^{-1}$) |
| 3.75 | Isua | 0.8 | 2.0 | 6.7 |
| 3.75 | Nuvvuagittuq | 0.1 | 0.2 | 0.7 |
| 3.25 | Moodies | 0.6 | 1.5 | 4.9 |
| 3.15 | Jharkhand-Orissa | 0.7 | 1.6 | 5.5 |
| 2.96 | Mozaan; Witwatersrand | 47.3 | 118.3 | 394.3 |
| 2.9 | Itilliarsuk | 0.1 | 0.2 | 0.5 |
| 2.85 | Slave Craton | 14.1 | 35.1 | 117.1 |
| 2.75 | Helen (Ontario) | 0.7 | 1.8 | 6.1 |
| 2.74 | Temagami | 7.3 | 18.3 | 60.9 |
| 2.74 | Carajás | 147.5 | 368.7 | 1228.9 |
| 2.72 | Hunter Mine Group | 0.7 | 1.8 | 6.1 |
| 2.72 | Bababudan Group | 7.3 | 18.3 | 60.9 |
| 2.7 | Anshan IF | 80.3 | 182.6 | 608.5 |
| 2.7 | Chitradurga Group | 7.3 | 18.3 | 60.9 |
| 2.7 | Tati | 15.0 | 37.5 | 125.0 |
| 2.62 | Slave Craton | 13.1 | 32.9 | 109.5 |
| 2.6 | Krivoy Rog IF | | | |
| 2.6 | Marra Bamba | 109.5 | 273.8 | 912.8 |
| 2.56 | Benchmark | 0.7 | 1.8 | 6.1 |
| 2.5 | Mt. Sylvia | 109.5 | 273.8 | 912.8 |
| 2.5 | Hutchison Group | 11.0 | 27.4 | 91.3 |
| 1.4 | Xiamaling | 12.8 | 31.9 | 106.5 |
| 1.36 | Jiamiao | 5.1 | 12.8 | 42.6 |
| 0.53 | Taxkorgan IF | 0.5 | 1.3 | 4.3 |

[1]Assuming a deposition rate of 6 m.Myr$^{-1}$
[2]Assuming a deposition rate of 15 m.Myr$^{-1}$
[3]Assuming a deposition rate of 50 m.Myr$^{-1}$

An alternative approach to calculate the Fe sink flux out of the oceans associated with IFs is to model the areal extent of IF deposition through time, which we can assume was likely restricted to the continental shelf. IFs were deposited below the storm wave base (~>150 m depth) but above the Ca carbonate compensation depth (CCD). If pCO$_2$ was high and the temperature of seawater was slightly above 0 °C, the Ca CCD could have been shallower than at present (Hakim et al., 2023), but ascribing a value is difficult. Given that the CCD is very sensitive to those parameters, this is of little help to constrain the maximum depth of IF deposition. We follow Trendall (2002) and assume that IF deposition extends to the edge of the continental slope, corresponding to a bathymetry of less than ~400 m. The rationale for focusing on continental shelf is that this is where biological productivity would be elevated so oxygen oases could develop, and iron in the photic zone could be efficiently shuttled to sediment. In open oceans, sinking iron(oxyhydr)oxide particles could be recycled in the water column by DIR. On modern Earth, the surface between isobaths -200 and -400 m represents a total area of 5,800,000 km$^2$. This environment would be susceptible to deposition of IF found at the edge of continents and it is reasonable to assume that it scales with coastal length taken at an appropriate scale. We have no direct constraints on coastal length through time, so we aim instead to find some scaling with the surface area of emerged lands. In a planar geometry, the 200-400 m depth area should scale linearly with coastal length, which itself should scale as the square root of the emerged lands. The total surface area of emerged lands at present is ~149,000,000 km$^2$. We thus expect, as a first-order scaling:

$S_{200-400} = 5,800,000/\sqrt{149,000,000} S_{Emerged(t)} = 475 S_{Emerged}$. In a spherical geometry, the scaling can become more complex, but this can be evaluated using a cap geometry for the emerged land. If we consider a cap of surface $A$ and circumference $C$ on a sphere of radius $R$, we expect the following relationships,

$$C = \sqrt{4\pi A - \frac{A^2}{R^2}}.$$

which has the limit $C^2 = 4\pi A$ for $R \to +\infty$, as is expected for a planar geometry. For the modern Earth, the equivalent cap circumference of emerged lands given by the first equation is ~36,406 km. If we write $S_{200-400} = wC$, where $w$ is the effective width of the area bracketed between 200-400 m bathymetry, we obtain $w = 159$ km. This is larger than the width of the region bracketed between the 200 and 400 m isobaths on most continental shelves because a significant fraction of this surface is in relatively flat epicontinental seas with broad extensions (Fig. 13). This characteristic length matches approximately the length scale of IF deposits, indicating that it is indeed a reasonable setting for IF deposition.

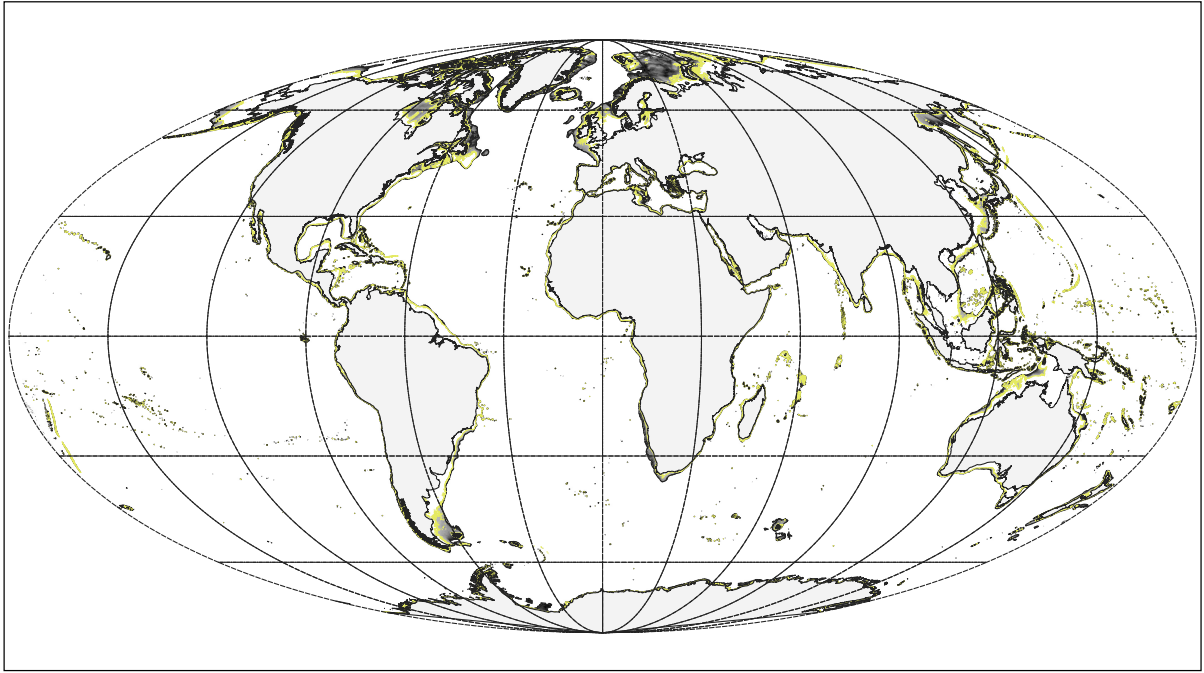

**Fig. 13**. World map showing the regions between 100 and 400 m isobaths highlighted in yellow.

We are interested in evaluating how the surface area bracketed between the 200-400 isobaths could have evolved through time as the continents grew. We can use the equation above to calculate $S_{200-400}(t)$,

$$S_{200-400}(t) = w\sqrt{4\pi A(t) - \frac{A(t)^2}{R^2}}.$$

This relationship is depicted in Fig. 14 (black dashed curve). To a reasonable extent, we can approximate this relationship by $S_{200-400}(t) \approx 511\sqrt{A(t)}$ with $S_{200-400}(t)$ and $A(t)$ both in km².

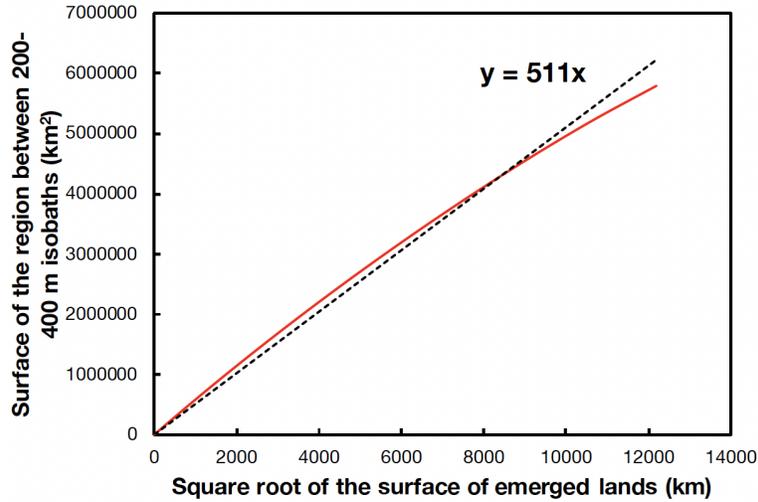

**Fig. 14.** Expected relationship between the surface area covered by seafloor between the 200 and 400 m isobaths (y-axis) and the square root of the total surface of emerged land (x-axis). The red line shows the expectation for a spherical cap geometry while the black dashed line is an approximation.

We further assume that continental surface area scales with continental volume, which grew linearly with time (McCulloch and Bennett, 1994) to its modern day value so we have $A(t) = 149{,}000{,}000 t/4.5 = 33{,}000{,}000 t$. This relationship is shown in Fig. 15, together with estimated surface area of emerged land over the past 600 Myr based on paleogeographic data. Combining the equations given above, the maximum areal extent of IFs (in km²) through time (in Gyr) would have been $S_{200-400}(t) = \alpha\sqrt{t}$ with $\alpha = 2{,}900{,}000$ km².Gyr$^{-0.5}$ (Fig. 15).

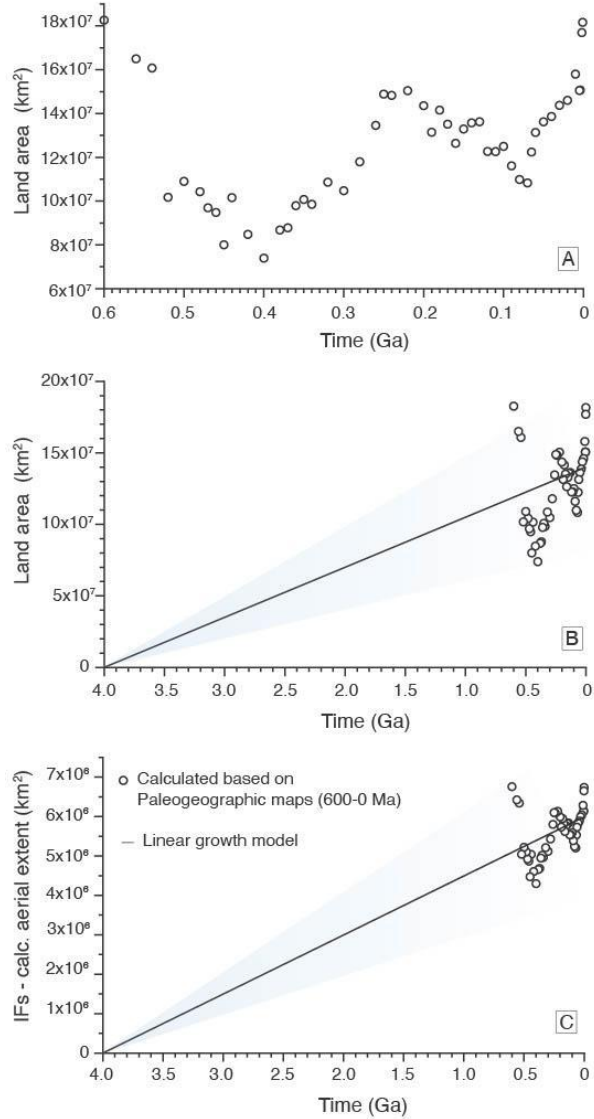

**Fig. 15.** Estimated land area and aerial extent of iron formation deposition through time (see text for details; a linear continental growth model is assumed, after McCulloch and Bennett (1994). The data points for comparison are estimates based on paleogeographic maps from Scotese (2016).

All the environments where IFs could be deposited would not necessarily host deposition of these sediments as other factors would surely play some role, such as chronological association with LIPs or spatial proximity to an upwelling region. To account for this, we add a 'fudge' factor $\lambda$ that can take its values between 0 (no IF deposition) to 1 (bathymetry is the only relevant consideration). We therefore have for the rate of Fe deposition in IFs,

$\phi_{Fe-IF} = \lambda S_{200-400}(t)\, \rho[Fe]_i = \lambda\, \alpha\, \sqrt{t}\, r_i \rho[Fe]_i.$

Assuming $\lambda = 1$, we calculate values of $\phi_{Fe-IF}$ up to 432320 Gg.yr$^{-1}$ during Stage #1. A more plausible assumption is that IF deposition only occurs in regions that see upwelling of $Fe^{2+}$-rich deep waters that boost biological productivity and photosynthesis. At present, the most prominent upwelling systems are found along the coasts of California, Peru, northwest Africa, and the west coast of southern Africa. They represent only a few percent of the world's total ocean area. Given the large uncertainty on the extent of upwelling regions during the Archean, and our interest in order of magnitude estimates, we set $\lambda$ to 0.1 for upwelling-limited IF deposition. Doing so, we obtain a value of $\phi_{Fe-IF} = 43232$ Gg.yr$^{-1}$ for the Archean (calculated at 3.0 Ga) by possibly overestimating IF depositional rates to 100 m.Myr$^{-1}$ (Konhauser et al., 2018; Lantink et al., 2022). This flux is significantly larger than the empirical rate derived from IF occurrences (e.g., average of $102 \pm 139$ Gg.yr$^{-1}$), and those two values represent reasonable upper and lower limits on the rate of IF deposition in the Archean.

**5.2. Iron silicates**

Iron silicate minerals that formed through chemical interaction with SiO$_{2(aq)}$ and dissolved organic carbon and/or through chemical weathering could have had a strong control on Fe concentration and electron availability for early anoxygenic phototrophs in early oceans (e.g., Holland, 1984; Holland et al., 1986). However, diagenesis and metamorphism obscure its quantification due to conversion of primary minerals to more stable minerals.

The Fe silicate flux has been strongly affected by the availability of dissolved silica in the ocean, and the marine silica sink. Today, siliceous skeleton production (mostly by diatoms with a minor contribution of radiolarian and sponges) leads to severe undersaturation with respect to solid silica phases (Treguer and De La Rocha, 2013) and negligeable precipitation and/or transformation of Fe-bearing authigenic clays (Isson and Planavsky, 2018). Considerable uncertainties exist on the magnitude of the modern reverse weathering flux, and modern estimates mostly focus on the Si budget (Rahman et al., 2017; Treguer and De La Rocha, 2013). Reverse weathering has been estimated to represent less than 10% of the total dissolved silica export to the modern ocean (3100 Gg Si.yr$^{-1}$). From this rough estimate, the removal of Fe from seawater and/or sediment porewater through reverse weathering can be estimated using the transformation reactions of pre-existing detrital clay minerals into the commonly found authigenic clays in the sedimentary record (Fe$^{2+}$:Si ratio from 0.16 to 2; Isson and Planavsky, 2018). By doing so, the modern Fe silicate flux amounts to 500 to 6000 Gg.yr$^{-1}$.

The absence of a biological silica sink in the Precambrian (stages #3 and #1) would have allowed the high dissolved SiO$_2$ concentration necessary for abiotic reactions in the water column and in sediment pore-waters (Tosca et al., 2019). Our current understanding of Fe-silicate formation relies on oversaturated Fe and SiO$_{2(aq)}$ laboratory experiments mimicking the Precambrian ocean (Halevy et al., 2017; Hinz et al., 2021; Tosca et al., 2016). Tosca et al. (2016) presented experimental evidence showing rapid nucleation of hydrous Fe$^{2+}$−`silicates, which eventually aggregate to form a precursor to the mineral greenalite at pH ≥ 7.5. More recently,` Hinz et al. (2021) documented greenalite formation at lower pH (down to 6.5) and demonstrated that the presence of Fe$^{3+}$ enhanced the greenalite precipitation. In the Archean ocean, greenalite precursors could have deposited in two settings: near hydrothermal vents and in coastal upwelling zones. In the absence of aqueous sulfide in the water column (the presence of sulfide would have rapidly depleted Fe$^{2+}$; Poulton et al., 2004), hydrothermal plumes were likely to meet all the required conditions (i.e., available Fe$^{2+}$, SiO$_{2(aq)}$, and pH between 6.5 to 8.3) for abundant greenalite precipitation (Rasmussen et al., 2021). Recently, the results of reaction path calculations investigating the mixing of high Fe/H$_2$S hydrothermal effluent and anoxic, sulfate-free Archean-analog seawater indicated that pyrite, greenalite, and/or siderite (depending on temperature kinetic inhibition; Greenberg and Tomson, 1992; Schoonen and Barnes, 1991) are thermodynamically favored across a wide parameter range (Tosca and Tutolo, 2023). Upon venting and mixing with ambient seawater, the above mineral assemblage might be responsible for 95 to 99% removal of the dissolved hydrothermal Fe in a sulfate-poor anoxic ocean (Tosca and Tutolo, 2023).

Halevy et al. (2017) observed the formation of mixed-valence, Fe(II,III) green rust upon partial Fe oxidation under low concentrations of silica (100-250 µM) at pH 7, transforming into three-layer Fe-silicate at low temperature (22 and 50 °C) in a few days to months. Despite being restricted to the upper ocean (near the Fe-redoxcline), Halevy et al. (2017) calculated that green rust may have been a global Fe sink, corresponding to between 60 and 100% of the global Fe-flux before the GOE (Stage #1). These authors suggested that green rust might have been an Fe-sink over the entire duration of stage #3. We note that these estimates only account for the above described oxidative pathway, although, more recently, Tosca et al. (2019) described a strictly anoxic pathway for green rust formation, which might increase the overall shuttle of Fe in stages #1 and #3.

In the Archean, iron silicates have been found associated with iron formations, often in the form of nanometer-scale crystals of greenalite in diagenetic chert nodules (Fig. 16, Rasmussen et al., 2015; Muhling and Rasmussen, 2020). Rasmussen and coworkers have made the case that these nanocrystals could be remnants of primary iron-bearing silicate gels that precipitated in Archean oceans, while the abundant iron oxides found in association were interpreted to be the products of post-sedimentary oxidation and separation into Si-rich and Fe-rich layers. This idea has not gained widespread acceptance as the extent of iron oxidation would be difficult to achieve after lithification (Robbins et al. 2019) and some geochemical proxies such as Fe/Mn-$\delta^{56}$Fe correlations point to some role for $Fe^{2+}$ oxidation in seawater (Thibon et al. 2019; Heard et al. 2022), although the equilibrium Fe isotopic fractionation associated with greenalite precipitation itself is still equivocal (Heard et al., 2023). If the interpretation of Rasmussen and coworkers is correct, then the IF-associated Fe oxide sink described in Sect. 5.1 would need to be labeled as a silicate sink, while the calculated Fe fluxes would still be relevant. In addition to iron formations, iron silicate may have been removed in a more diffuse manner, in hydrothermal plumes (Tosca and Tutolo, 2023) or from seawater in areas with high Si and Fe contents that experienced partial oxidation (Hinz et al. 2021). Given these large uncertainties on the location and conditions of iron silicate removal, and given the absence of clear geological archives of these deposits, we have decided to leave this sink as an unknown that will be adjusted if the mass balance calls for it.

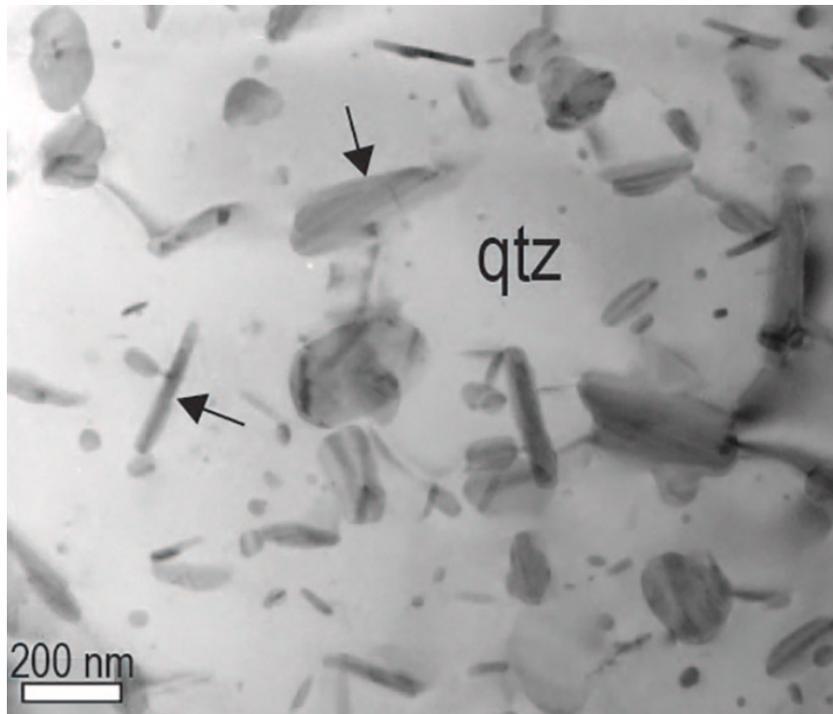

**Fig. 16.** Bright-field transmission electron microscope image of greenalite nanoparticles (arrows) in fine grain quartz (qtz) from Bee Gorge Member, Wittenoom Formation (Rasmussen et al., 2021).

**5.3. Iron carbonates**

In some environments, Fe can precipitate as iron carbonates, mostly siderite, ankerite, or ferroan-rich dolomite. These minerals contain up to few tens of weight percent of Fe. Today, most iron-bearing carbonates are formed in peculiar meromictic lakes such as lake Towuti (Indonesia; Vuillemin et al., 2019) or Otter Lake (USA; Wittkop et al., 2014), both of which are ferruginous and stratified. In those environments, Fe-rich carbonate (mostly siderite) is formed diagenetically following sedimentary organic matter degradation and resultant high DICin porewaters. Vuillemin et al. (2019) found evidence that siderite growth proceeds through aging of green rust, consistent with the experimental results of Halevy et al. (2017). However, low concentration of $Fe^{2+}$ in the modern global ocean limits the Fe-rich carbonate sink to trace incorporation into marine calcifiers (bivalves, corals and planktonic organisms; e.g., Emiliani, 1955), which therefore represent a negligible iron flux.

The geological record of sedimentary rocks preserves some evidence of an active carbonate cycle since early in Earth's history. The past carbonate factory was very different from the modern one dominated by biomineralization processes as it is thought to have functioned via inorganic precipitation within the water column or at the seafloor (e.g., precipitating aragonite fans and thick calcite encrustations; Kah and Bartley, 2021; Sumner and Grotzinger, 2004). Using the widespread occurrence of dolomite and calcite instead of siderite in the geological record, Holland (1984) proposed that the Archean oceans should have been poised between calcite and siderite saturation. Jiang and Tosca (2019) experimentally re-evaluated Fe carbonate precipitation in the presence or absence of dissolved silica using Archean-like anoxic seawater. Their results showed that direct (spontaneous) precipitation from Archean-like anoxic and low-sulfate seawater might result in the precipitation of Fe-rich carbonate such as chukanovite or siderite while in the presence of dissolved silica, spontaneous Fe-rich carbonate precipitation is kinetically inhibited. Instead, the main product appears to be Fe silicate (greenalite). Similarly, there is evidence supporting an early diagenetic origin of siderite in the majority of IFs preserved in the geologic record (Raiswell et al., 2011; Rasmussen and Muhling, 2018). As such, we have neglected the Fe-rich carbonate sink, in part because it is likely already counted as a part of the iron oxide IF sink.

### 5.4 Pyrite

Pyrite is the most abundant sulfide mineral on Earth and can be found ubiquitously in sedimentary settings (Fig. 17). Over a long timescale, the burial of pyrite represents an indirect net source of oxygen to the ocean-atmosphere sytem and is, therefore, an important factor regulating Earth's surface redox state (e.g., Berner, 1984). Pyrite is formed through abiotic and biological reactions between $Fe(II)_{aq}$ and sulfide ($H_2S$ and $HS^-$). The Fe cycle is closely coupled to oxygen, carbon, and sulfur cycles. It interacts with the carbon cycle through the process of microbial DIR (Kappler et al., 2021). Alternatively, $Fe(II)_{aq}$ may also be abiotically released to pore waters through sulfide-mediated reduction of Fe(III) (oxyhydr)oxides (Peiffer et al., 2015; Wan et al., 2017). In pore waters, and certain reducing water columns reviewed in Section 2, $Fe(II)_{aq}$ may react with sulfide ($H_2S$ and $HS^-$) to form sedimentary pyrite ($FeS_2$; Rickard and Luther, 2007).

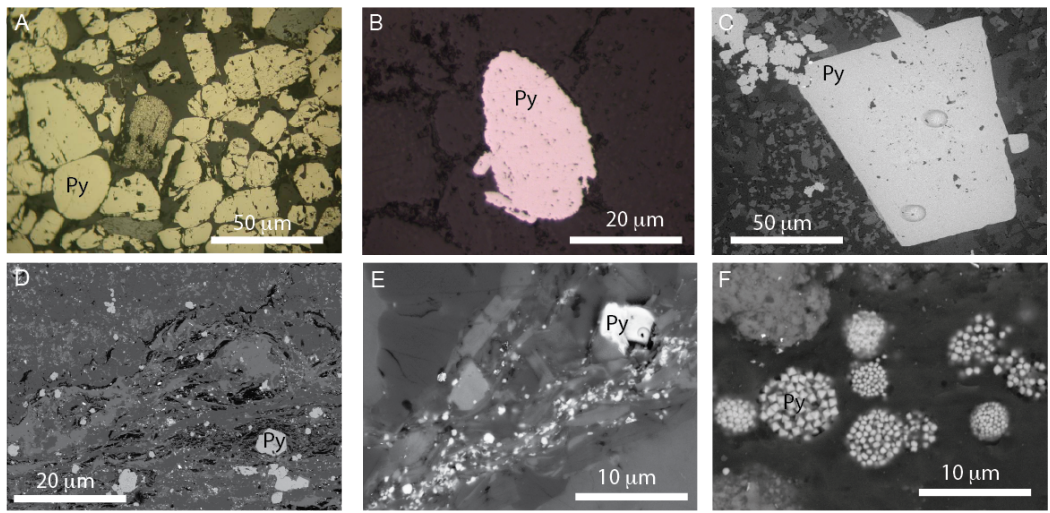

**Fig. 17.** Examples of sedimentary pyrite minerals of Archean age. A and B are detrital pyrite from the Witwatersrand and the Moodies Group, C is massive pyrite from Mendon chert, D and E are micropyrites from Buck Reef and Tumbiana, and F is framboidal pyrite from Atexcac lake.

In modern environments at near-neutral pH, pyrite forms at the expense of local Fe(III) mineral phases and OM following microbial sulfate reduction and resultant sulfide production, all of which explains why its distribution in marine sediment is heterogeneous. Estimating the marine Fe sink flux associated with pyrite burial is relatively straightforward as there is a negligible amount of pyrite re-oxidation, thus, all pyrite preserved in sedimentary rock can be used to determine the Fe flux. To do so, two approaches can be taken. One is to study sediment cores or sedimentary rocks; the other is to use geochemical models. In doing so, the former can be viewed as an upper limit on the $Fe_{PYR}$ burial flux as it does not discriminate the origin of pyrite preserved in sedimentary rocks (i.e., syn- vs. post-depositional), whereas the latter represents a lower estimate.

The surface area over which marine sediment is accumulating today is difficult to estimate precisely because sedimentation rate can vary widely depending on location, water depth, and hydrodynamic conditions. However, a rough estimate can be made based on the total surface area of the world's ocean ($362.10^{12}$ m$^2$), an average sedimentation rate for the global ocean (0.1 mm yr$^{-1}$), average sediment density of 2650 g.m$^3$, and an average sediment porosity of 0.4 (e.g., Boudreau, 2000). Such a calculation yields an average mass of sediment deposited of 58 Tg.yr$^{-1}$. The total Fe concentration in modern marine sediment is ~3.2 wt% and about 10% of that Fe is preserved in the form of pyrite (Pasquier et al., 2022), corresponding to a present, pyrite burial Fe flux of ~185 Gg.yr$^{-1}$ (Table 2). We emphasize that this number is a rough calculation using statistics based on a limited number of observations (n~320) with a few locations at water depth below 1000 m (20% of the database), which tend to have lower $Fe_{PYR}$ values than shallower depositional environments. Several global biogeochemical models of C and S have shown that pyrite burial fluxes varied through time (e.g., Berner and Canfield, 1989). Using the same approach over the entire duration of stage #5 with the appropriate $Fe_{PYR}/Fe_T$ ratio yields a $Fe_{PYR}$ burial flux of 360 Gg.yr$^{-1}$ (Table 2). Assuming sediment deposition rates of 1 mm.yr$^{-1}$ and similar physical properties (density and porosity), the empirical data-based approach yields $Fe_{PYR}$ fluxes of 305 and 880 Gg.yr$^{-1}$ for stages #5 and #1, respectively (Table 2).

It is important to keep in mind that the Fe speciation compilation is mostly composed of shallow and siliciclastic marine settings, both of which are likely to contain more pyrite than basinal and carbonate lithologies. Consequently, our estimate likely represents an upper limit on the true modern $Fe_{PYR}$ flux. As mentioned above, it is hard to quantify how post-depositional processes such as diagenesis and metamorphism may have compromised quantification of $Fe_{PYR}$ from examination of sedimentary rocks. For example, fluid-rock interaction has been shown to result in late-stage precipitation of pyrite and pyrrhotite which may have led to spuriously high pyrite content (e.g., Slotznick et al., 2018; Slotznick et al., 2022).

**Table 2.** Pyrite burial flux at different stages.

| Stage | Age (Ga) | n | $Fe_T$ (wt.%) | $Fe_{PYR}$ (wt.%) | Fe flux (Gg.yr$^{-1}$) |
|---|---|---|---|---|---|
| #5 | 0 - 0.56 | 9599 | 3.04 | 0.63 | ≈ 360 |
| #3 | 0.8 – 1.85 | 2346 | 4.05 | 0.51 | ≈ 305 |
| | | | | | |
| | | | | | |
| #1 | > 2.5 | 458 | 4.78 | 1.27 | ≈ 880 |
| Modern | | 2990 | 3.18 | 0.31 | ≈ 185 |

Because the burial of pyrite is coupled to the S cycle, an alternative to the above is to estimate the pyrite burial ($f_{PYR}$) using a steady-state approach constrained by S isotope values. In doing so, it is important to keep in mind that $f_{PYR}$ critically depends on the S isotopic composition of the riverine input, which in modern environments is strongly impacted by dissolution of recent evaporite deposits (i.e., larger contribution from evaporitic sequences resulting in a lower $f_{PYR}$). After accounting for the above, Halevy et al. (2012) estimated that the long-term Phanerozoic (stage #5) $f_{PYR}$ to be between 70 and 90% of the riverine sulfate input ($3.10^{12}$ mol S.yr$^{-1}$), which corresponds to a Fe$_{PYR}$ burial rate between 60 and 75 Gg.yr$^{-1}$. This range is in a relatively good agreement with the Fe pyrite burial flux ranging between 47 and 55 Gg.yr$^{-1}$ required to satisfy the mass balance of the oxidized S species over the last 60 Ma (Rennie et al., 2018). Similarly, Fakhraee and Katsev (2019) `used a mass balance coupled to a 1D diagenetic model to reconstruct the S-cycle during the Proterozoic eon (≈stage #3) from which they estimated an Fe`$_{PYR}$ burial between 25 to 55 Gg.yr$^{-1}$. For stage #1, model estimates of Fe pyrite burial fluxes range from ~2.6 to 28 Gg.yr$^{-1}$ (e.g., Fakhraee and Katsev, 2019; Laakso and Schrag, 2017).

### 5.5 Sedimentary source of Fe to the oceans through time.

As outlined in Section 2, some $19 \times 10^{15}$ g.yr$^{-1}$ of suspended sediment is delivered to the oceans every year by modern global river discharge (Peucker-Ehrenbrink, 2009). Assuming a 3.5 wt% Fe content for the Phanerozoic upper continental crust (Ptáček et al., 2020), this sediment represents a flux of 665,000 Gg.yr$^{-1}$ of Fe, only a fraction of which can interact with the DIB cycle. In the modern, oxic ocean, a small fraction of sedimentary Fe is ultimately released to the oceans, in high-productivity areas where organic carbon loading establishes anoxic porewater conditions and enables microbial DIR of Fe(III) in minerals (Section 2). The 8400 Gg.yr$^{-1}$ modern sedimentary dissolved Fe flux estimated by Dale et al. (2015) is just 1.25 % of all sedimentary Fe delivery from the continents to the marine realm, reflecting the dominance of oxic conditions on the modern Earth that enable burial of Fe(III) minerals.

Under less oxygenated Precambrian atmospheric conditions, particularly in the Archean (Stage #1), less sedimentary Fe was delivered to the oceans in sediments in Fe(III) mineral phases utilizable for microbial DIR, but more Fe(II) should have been carried in rivers prior to sedimentation. Taking a similar approach to our treatment of the aeolian Fe flux above in Section 3.1 and assigning a 10-60 % solubility range to riverine particulate Fe, we estimate a 'sedimentary' Fe flux to the oceans of 66,000 to 400,000 Gg.yr$^{-1}$. In practical terms this Fe flux reflects a double counting of the riverine dissolved Fe flux calculated in Section 3.2, because under an anoxic atmosphere, Fe would have simply been released to solution instead of traveling with suspended particles during riverine transport. The fact that we thus predict a range of riverine Fe fluxes in Stages #1 and #3 that overlap with estimates in Section 3.2 using this independent approach supports the robustness of those calculations.

Accepting that the terrigenous sedimentary and dissolved fluxes of Fe to the oceans may have been somewhat interchangeable through geological history according to the redox conditions of continental weathering, the remaining pertinent, 'sedimentary' Fe source to the oceans through Earth history should have been the operation of microbial DIR in environments where Fe oxides were delivered to the seafloor. Specifically, microbial DIR may have recycled some fraction of Fe oxides formed during deposition of Precambrian Fe oxide-rich sediments such as IFs. Microbial DIR in IFs has long been a favored interpretation for the mixed valence ($Fe^{2.4+}$) nature of these deposits under the assumption of a purely ferric mineral precursor (Klein, 2005; Konhauser et al., 2005). Ferric Fe in IF mineral precursors deposited on an anoxic seafloor would have been favorable electron acceptors for the oxidation of organic carbon (Walker, 1984) that present-day IFs are notably deficient in (0.01-0.5 wt % $C_{org}$; Gole and Klein, 1981; Konhauser et al., 2017). Additionally, the oxidation of organic matter has been proposed to explain the negative $\delta^{13}C$ and positive $\delta^{56}Fe$ values of IF carbonates (Baur et al., 1985; Perry et al., 1973; Walker, 1984; Craddock and Dauphas, 2011; Heimann et al., 2010).

Microbial DIR in anoxic sediments modifies the Fe oxide sink flux, and therefore we can tie estimates of the DIR flux back to the oceans to our calculated oxide sink in Section 5.1. The presence of Fe(II) mineral phases in IFs under the assumption of a ferric mineral precursor suggests that only a fraction of $Fe^{2+}$ generated during DIR was transported back into the water column, with the rest being retained in diagenetic Fe(II) phases. However, the fraction of Fe recycled into seawater by DIR during IF deposition has proven extremely difficult to quantify (Konhauser et al., 2005), in part because it depends on the initial oxidation mechanisms that imply very different Fe:$C_{org}$ ratios, ranging from infinite (1:0) for purely abiotic UV photo-oxidation, to 4:1 for photoferrotrophy. Therefore the lower limit for the DIR flux is ~0 % of the IF deposition flux, in the case of either 1) little to no coupled burial of Fe and $C_{org}$, or 2) closed system DIR after burial that retained diagenetic Fe(II) in sediments. On the upper end, DIR should be most effective if Fe oxides and $C_{org}$ formed in close association with one another; as would occur if Fe(II) oxidation was directly mediated by photoferrotrophic bacteria. In this specific scenario, Konhauser et al. (2005) suggested that up to 70 % of Fe oxides precipitated in the Archean oceans could have ultimately been recycled in the water column as diagenetic Fe(II), based on a comparison of optimal photoferrotrophic productivity rates and the IF burial rates. In this case, the DIR flux back into the oceans could have been 7/3 times the IF deposition flux calculated above in Section 5.1. This upper limit could apply to not only Stage #1, but also Stage #3, where photoferrotrophic involvement in IF deposition has also been proposed (Canfield et al., 2018; Wang et al., 2022). These scalings indicate a range of Fe input fluxes from **0 to 13,600 Gg.yr$^{-1}$** in Stage #1 and from **0 to 347 Gg.yr$^{-1}$** in Stage #3.

## 6. Balancing the biogeochemical cycle of iron

| Table 3. Iron flux in and out of the oceans in Gg.yr$^{-1}$. Fluxes in and out are positive and negative. | | | | |
|---|---|---|---|---|
| | | Pre-GOE | Post-GOE/Pre-NOE | Post-NOE |
| | | Stage #1 | Stage #3 | Stage #5 |
| Source/Sink | Secti | 3.85-2.4 Ga | 2.1-0.8 Ga | 0.56 Ga-present |

|  |  | on |  |  |  |
|---|---|---|---|---|---|
| Aeolian dust | | 3.1 | 500 to 10,200 | 760 to 15,500 | 460 (70 to 2,400) |
| Rivers | | 3.2 | 61,000 to 182,000 | 4 to 370? | 145 (4 to 250) |
| Hydrothermal | | 4 | 30 to 57,000 (1 to 2280 if greenalite precipitates efficiently at vents) | 30 to 17,000 | 760 (280 to 1,240) |
| Sedimentary sources | | | 0 to 13,600 | 0 to 347 | 8400 |
| Sedimentary sinks | | 5 | | | |
| | Oxide | 5.1 | -70 to -43232 | -45 to -149? | ? |
| | Sulfide | 5.2 | -3 to -880 | -25 to -305? | -360 |
| Silicate | | | ? | ? | Negligible |
| Total sources | | | 61501 to 262,800 | | 8754 to 12290 |
| Total sinks | | | -73 to -44112 | | |
| Note. The fluxes with the greatest uncertainties are indicated with a question mark. Fluxes lacking constraints are indicated only with a question mark. | | | | | |

The inferred iron fluxes in and out of the oceans in the Archean and at present are compiled in Table 3 and depicted graphically in Fig. 18. As highlighted throughout this contribution, the Proterozoic eon is a period of greater uncertainty with regard to the iron biogeochemical cycle and we will therefore not discuss the balance of sources and sinks during that time.

We find that in the Archean, continental export likely represented a significant source of iron, but hydrothermal input could also have been significant depending on the amount of iron immobilized as carbonate in hydrothermally altered oceanic crust or through proximal scavenging in the hydrothermal plume. All sources together would have delivered **61,500 to 263,000 Gg.yr$^{-1}$** of dissolved iron to the oceans. Balancing this large iron source with inventoried sinks is not straightforward. Indeed, examination of IF occurrences gives an average iron deposition rate of **~100 Gg.yr$^{-1}$** and sums up to ~600 Gg.yr$^{-1}$ towards the end of the Archean (i.e. 2.74 to 2.7 Ga). Even in an optimistic scenario where IFs were deposited in upwelling regions on seafloor at a bathymetry between 200 m and 400 m depth, we calculate an iron sink of **~43000 Gg.yr$^{-1}$** at most (sulfides play a lesser role). There appears to be a significant imbalance between sources and identifiable sinks for dissolved iron in the Archean. This imbalance can be solved in two ways that both involve dispersed removal of iron into extended sinks. The missing sink could have been iron oxides removed from the photic zone and sinking into deep water basins that were recycled either to the mantle or in orogens and are missing from the recognizable, low-metamorphic grade sedimentary record. Another possibility is that iron was removed abiotically as silicate (Tosca et al. 2016; Hinz et al. 2021) or carbonate (Holland, 1984; Grotzinger and Kasting, 1993; Canfield, 2005; Raiswell et al., 2011) minerals mediated through iron supersaturation for these minerals. Precipitation of these minerals is affected by pH, silica and carbonate activities, temperature, and iron oxidation; and requires overcoming kinetic barriers, so it is impossible to tell which of these two mineral precipitates could have been the missing sink needed to balance iron sources in the Archean.

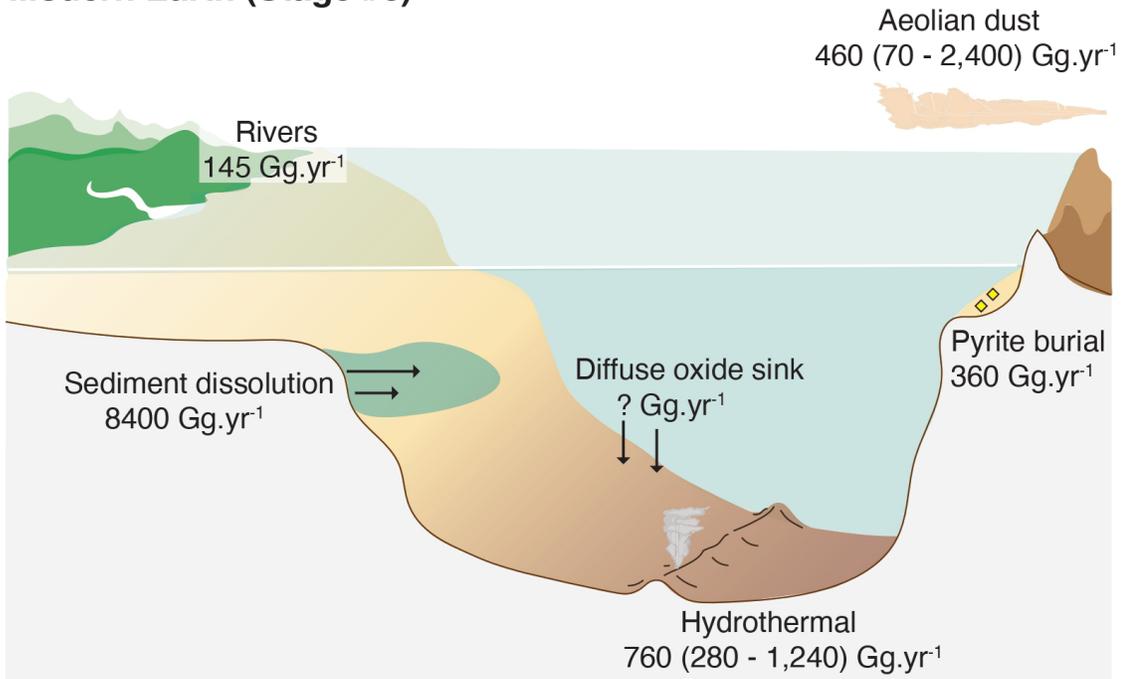
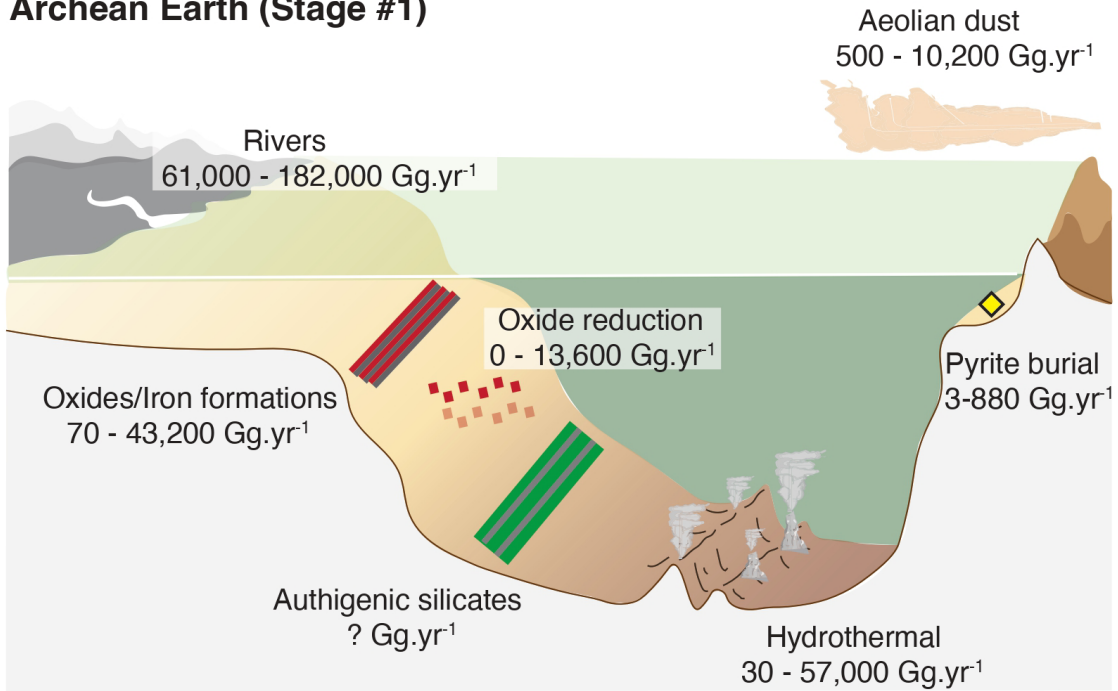

**Fig. 18.** Overview illustration of iron fluxes in the modern (stage #5) and Archean (stage #1) Earth.

# 7. The residence time of dissolved iron in Archean oceans

Knowing the dissolved Fe inventory of the ocean, and either the flux in or out (F), we can calculate the residence time of dissolved iron; $t_{res}=[Fe]_{SW}M_{SW}/F$, where $[Fe]_{SW}$ is the average seawater concentration of iron, $M_{SW}$ is the mass of the oceans ($1.4\times10^{21}$ kg). The average Fe concentration in the modern ocean is $0.58 \pm 0.14$ nM (Tagliabue et al., 2016). This corresponds to a total dissolved Fe budget of $45350 \pm 10950$ Gg. Using the flux values compiled in Table 3, we calculate a dissolved Fe residence time of around 5 yr. For comparison, Tagliabue et al. reported Fe residence time in the modern ocean ranging between 4 and 626 yr that overlap with the value estimated here, but is higher overall. The main source of uncertainty in this calculation is the sedimentary flux of iron, which is difficult to quantify and may be locally balanced by near-seafloor scavenging processes that prevent major Fe delivery to the upper ocean (Dale et al., 2015).

As discussed above, sedimentary sinks of dissolved iron in the Archean are poorly constrained because of difficulties in reading the incomplete and presumably biased geological record. In calculating the residence time of iron in the Archean, it therefore makes more sense to rely on input fluxes. These would have been dominated by rivers and hydrothermal sources. The total input flux of Fe during that time would have been between 60,000 and 240,000 Gg/yr (Table 3). These values are relatively well constrained as the riverine flux is dominant and the Fe flux is calculated only assuming that on a global scale, $CO_2$ degassing is balanced by seafloor and continental weathering. The main uncertainty in calculating the residence time of Fe is in estimating the dissolved Fe inventory of seawater. Estimates of dissolved iron concentration in the Archean range between 20 $\mu$M and 2 mM, based on consideration of the saturation of siderite and greenalite (Holland 1984; Canfield 2005; Jiang and Tosca 2019). These estimates correspond to total seawater inventories of dissolved iron of $1.5 \times 10^9$ to $1.5 \times 10^{11}$ Gg. We thus calculate a residence time for dissolved iron in the Archean ocean of 6 kyr ($1.5 \times 10^9/240,000$) to 3 Myr ($1.5 \times 10^{11}/60,000$).

Johnson et al. (2008) calculated the residence time of iron in the Archean ocean using previously published data on iron content and flux, and obtained values between 10 and 240 kyr. The dissolved iron concentrations that they considered range between 36 and 895 $\mu$M (Ewers, 1983; Sumner, 1997; Canfield, 2005), so not very different from the range of 20 to 2000 $\mu$M considered above. They adopted an iron flux from Canfield and Raiswell (1999) of 280,000 Gg.yr$^{-1}$, which is close to the upper limit of the flux that we adopted (240,000 Gg.yr$^{-1}$). An important difference between the calculation of Johnson et al. (2008) and ours is in the lower limit of the iron flux value that we consider of 60,000 Gg.yr$^{-1}$. This lower value is calculated assuming that (*i*) hydrothermal iron is immobilized in or near vent site by carbonation of the oceanic crust or by rapid precipitation of iron-silicates in the hydrothermal plume, and (*ii*) the rate of $CO_2$ degassing was similar in the Archean to what it is at present. Under those conditions, we can reach a residence time for iron in the Archean oceans of 3 Myr.

Thibon et al. (2019) used an empirical approach to constrain the residence time of iron by using a series of IFs that showed exponential evolution in their Fe isotopic composition consistent with the e-folding response of a system to an instantaneous disturbance. They thus obtained Fe residence time between ~2.5 and 2.4 Ga of 200 kyr to 2.3 Myr. These residence times largely overlap with the ones that we calculated above.

Overall, our estimate of the residence time of iron in the Archean oceans using the fluxes calculated above encompasses other estimates indicating that in the absence of oxygen in the atmosphere, it would have been much longer than today. The mixing timescale of the modern ocean is around 1 kyr. Ocean mixing is influenced by many parameters that could have been different in the Archean compared to what they are at present. The proximity of the Moon to the Earth, faster rotation rate, different continental configuration, insolation, ocean salinity, and several other factors were different in the Archean that could have affected ocean mixing timescale. Scaling arguments (Chen et al., 2021) and detailed modeling (Liu et al., 2023) show however that changes in these parameters did not modify the ocean mixing timescale beyond a factor of ~2 up or down relative to the present time (Liu et al., 2023). The residence time of iron in the Archean ocean of 6 kyr to 3 Myr would have been significantly longer than the ocean mixing timescale during that time, meaning that to first order, the concentration and isotopic composition of iron in open seawater must have been homogeneous. Our analysis also shows that continental export (primarily rivers) would have been the main contributor to dissolved iron (~50-100%) with hydrothermal sources representing ~0-50% (Table 3). These conclusions apparently contradict geochemical evidence indicating that (*i*) iron in banded iron formations had a significant mantle hydrothermal contribution (Derry and Jacobsen, 1990; Hu et al., 2020; Jacobsen and Pimentel-Klose, 1988; Wang et al., 2016) and (*ii*) oxidation of iron likely produced water masses with fractionated Fe/Mn and $\delta^{56}$Fe (Thibon et al. 2019; Heard et al. 2022). However, the calculated residence time pertains mainly to large-scale oceanic conditions and may not reflect the variability in local depositional environments, such as areas where deep-sea upwelling currents transported hydrothermal iron to photic zones abundant with biotic or abiotic oxidants.

## 8. Conclusion

The evolution of seawater-dissolved iron through Earth's history was complex, reflecting the multifaceted biogeochemical behavior of this element, which is involved in redox reactions and is an important bionutrient. Our research focused on periods marked by relative stability in Earth's surface redox evolution, but even within these periods, significant uncertainties persist about the origins and fate of dissolved iron. In the Archean eon, riverine export of iron from continents may have been highly significant, possibly dominating the influx of iron into the oceans. Based on our estimates of iron input to the Archean oceans and prior data on dissolved iron concentrations, we calculate a residence time ranging from 6,000 years to 3 million years. This notably exceeds the inferred oceanic mixing timescale at that time, which was 500 to 2,000 years, suggesting that dissolved iron likely had a uniform concentration and isotopic composition across different oceanic basins. For comparison, the residence time of iron in modern oceans is only 5 years.

Our analysis reveals that Archean iron sources cannot be fully accounted for by clearly identifiable sinks. Even when accounting for iron deposition on continental shelves at depths of 200 to 400 meters in regions where upwelling may have occurred—and considering the potential increase in continental landmass over time—iron formations still appear to be an inadequate sink to offset the total sources. This indicates that much of the iron was removed from the oceans via a cryptic sink, possibly as carbonate or silicate, extending into regions of the oceans not preserved in the rock record. This sink may be under-represented because it was subsequently subducted into the mantle.

**Acknowledgements.** We thank Sasha Turchyn for her editorial handling, and Carrie Soderman for her careful and constructive review of the manuscript. This review is based upon work supported by the U.S. Department of Energy, Office of Science, Office of Basic Energy Sciences, Geosciences program under Award Number DE-SC0022451 and NASA grant 80NSSC20K1409 (Habitable Worlds).